\documentclass[amssymb,amsmath,showpacs,prb,twocolumn,superscriptaddress,floatfix]{revtex4}
\usepackage{graphicx}
\usepackage{dcolumn}% Align table columns on decimal point
\usepackage{bm}% bold math
\usepackage[up]{subfigure}

\newcommand{\uB}{\ensuremath{\mu_{\text{B}}}}

\begin{document}

%\preprint{}

\title{Magnetic structure of small Fe, Mn, and Cr clusters supported on
   Cu(111): Noncollinear first-principles calculations}

\author{Anders Bergman}
\altaffiliation[Current address:]{
 D\'epartement de Recherche Fondamentale sur la Mati\`ere Condens\'ee,
SP2M/LSim, CEA-Grenoble, 38054 Grenoble Cedex 9, France}
\affiliation{Department of Physics, Uppsala University, Box 530 Sweden}
\author{Lars Nordstr\"om}
\affiliation{Department of Physics, Uppsala University, Box 530 Sweden}
\author{Angela Burlamaqui Klautau}
\affiliation{Departamento de F\'\i sica, Universidade
Federal do Par\'a, Bel\'em,
 PA, Brazil}
 \author{Sonia Frota-Pess\^oa}
\affiliation{Instituto de Fisica, Universidade de S\~ao Paulo, CP 66318, S\~ao Paulo, SP, Brazil}
\author{Olle Eriksson}
\email{Olle.Eriksson@fysik.uu.se}
\affiliation{Department of Physics, Uppsala University, Box 530 Sweden}
\date{\today}

\newcommand{\tr}{\operatorname{tr}}

\begin{abstract}
The magnetic structures of small clusters of Fe, Mn, and Cr supported
on a Cu(111) surface have been studied with non-collinear first
principles theory. Different geometries such as triangles, pyramids
and wires are considered and the cluster sizes have been varied
between two to ten atoms. The calculations have been performed using a
real space linear muffin-tin orbital method (RS-LMTO-ASA). The Fe
clusters are found to order ferromagnetically regardless of the
cluster geometry. For Mn and Cr clusters, antiferromagnetic exchange
interactions between nearest-neighbours are found to cause collinear
antiferromagnetic ordering when the geometry allows it. If the
antiferromagnetism is frustrated by the cluster geometry non-collinear
ordering is found. A comparison between the calculated structures and
ground states obtained from simplified Heisenberg Hamiltonians show
that the exchange interaction varies for different atoms in the
clusters as a result of the different local structure.
\end{abstract}

% 75.75.+a 	Magnetic properties of nanostructures
% 73.22.-f 	Electronic structure of nanoscale materials: clusters, nanoparticles, nanotubes, and nanocrystals
% 75.10.-b 	General theory and models of magnetic ordering 
\maketitle

\section{Introduction}
The remarkable progress of experimental methodologies with atomic
resolution, such as the scanning electron microscopy\cite{Binnig}
(STM), has paved the way for studies of nanoscale magnetic materials
such as ad-atoms, clusters and wires deposited on surfaces. As a
result of the reduced dimensions and symmetries for such systems,
magnetic behaviour that differs from bulk materials can be
found\cite{Himpsel,Gambardella2001}. This attracts interest not only
for the novel physics that can occur in these systems but also for the
possibility to tailor the electronic and magnetic properties by
changing the structure and the local environment of the systems.  
\par
Studies of systems consisting of only a few atoms can give valuable
information on how the magnetic structure evolves from single atoms
towards the bulk behaviour. Fe, Mn, and Cr are all known to exhibit
interesting magnetic behaviour. While being ferromagnetic in the bcc
phase, Fe in the fcc phase has been found to exhibit a spin-spiral
structure when synthesized as precipitates in a Cu
matrix\cite{Tsunoda89} and calculations show that the magnetic
structure is strongly dependent on the lattice
parameter\cite{Sjostedt02}. Cr has in bulk an incommensurate
antiferromagnetic spin density wave\cite{Fawcett88} which can be tuned
by creating superlattices with ferromagnetic or paramagnetic layers
and varying the interface roughness and layer
thickness\cite{Bodeker99}. When deposited on stepped surfaces, Cr can
be found to have non-collinear ordering\cite{Robles03}. Bulk Mn
exhibits perhaps the most intriguing magnetic structure of all
elements with a unit cell containing 58 atoms\cite{Bradley27} with a
complex non-collinear antiferromagnetic magnetic
structure.\cite{Hobbs03} Recently, several Mn based compounds, where
the magnetic ordering is non-collinear due to geometric frustration
between the magnetic moments of Mn atoms, have been
studied\cite{teriksson04,teriksson05} experimentally as well as
theoretically. 
\par
Free clusters of Fe, Mn, and Cr have been studied both experimentally
and theoretically. Stern-Gerlach measurements on Fe\cite{Billas93}
clusters show ferromagnetic behaviour while Mn
clusters\cite{Knickelbein01} and Cr clusters\cite{Douglass92} show
varying small net deflections, which can be interpreted as the result
of antiferromagnetic or even non-collinear magnetic
configurations. Calculations have shown that small Mn and Cr clusters
can exhibit non-collinear magnetic
ordering\cite{Kohl99,Morisato05,Longo05} in agreement with the
Stern-Gerlach experiments.  
\par
Supported transition metal clusters have also been studied extensively
where a majority of the studies have been theoretical. Among the
experimental studies, most work has been done on Fe, where Fe clusters
deposited on a Ni surface have been found to be ferromagnetic with
oscillating magnitude of the orbital moments\cite{Lau02} and  Fe
clusters supported on a graphite surface have been found to exhibit
enhanced spin and orbital moments compared to
bulk\cite{Binns2001}. Among the theoretical studies, the reported
calculations have mostly only considered collinear magnetization
densities. Monoatomic wires of Fe on Cu(111) and Cu(001) show
ferromagnetic behaviour with a strong magnetic
anisotropy\cite{Spisak02,Lazarovits03}. Small Mn clusters on Ag(001)
have been found to exhibit magnetic bi-stability
\cite{Stepanyuk97,Stepanyuk97_2} which is also the case for mixed
clusters of FeMn and FeCr that have been found to have both ferro- and
antiferromagnetic solutions close in energy\cite{Stepanyuk98}. Early
model calculations of supported equilateral triangular transition
metal clusters have shown that non-collinear ordering can be obtained
from the frustration due to antiferromagnetic interactions between the
cluster atoms.\cite{Uzdin1999,Uzdin2001} Recent studies have found that
small clusters of Mn and Cr become non-collinear when deposited on
Ni(001)\cite{Lounis05} and Fe(001)\cite{Robles,Lounis06} surfaces due
to competing exchange interactions between the cluster atoms and the surface\cite{Lounis05}. 
\par
In a previous paper\cite{Bergman06} we reported on non-collinear
magnetic ordering for a selection of small Mn clusters supported on a
Cu surface. In this paper we expand these results and present
theoretical results concerning the magnetic ordering and interactions
for Fe, Mn, and Cr clusters deposited on a Cu(111) surface. The
calculations have been performed using the RS-LMTO-ASA method that is
a first principles order-N method which has recently been extended to
the treatment of non-collinear magnetism\cite{Bergman06}. 
\section{Method}
The RS-LMTO-ASA method is based on the LMTO-ASA
technique\cite{andersen75} and the Haydock recursion
method\cite{haydock92}. The LMTO-ASA formalism provides an efficient,
parameter-free, basis set for treating close packed metallic systems
and the recursion method gives the ability to treat problems where
translational symmetry is absent and does also convey order-N scaling
with respect to the number of nonequivalent atoms in the system.  
The recursion method does not directly solve the eigenvalue problem as
formulated in the DFT, but allows one to calculate the local density
of states (LDOS) for the orbitals of the atoms in the selected system. The
RS-LMTO-ASA method has successfully been used for a wide range of
problems including bulk systems, multilayers, embedded impurities and
clusters and clusters on surfaces. Earlier and more detailed
descriptions of the collinear implementation of the RS-LMTO-ASA can be
found elsewhere\cite{Frota92,Klautau05}.   
\par
In the local spin density approximation (LSDA)\cite{barth72}, the
electron density is expressed through a 2x2 density matrix $\rho$
which can be expressed in terms of the non-magnetic charge density $n$
and the magnetization density $\bm{m}$ as
$\rho=(n\mathcal{I}+\bm{m}\cdot\bm{\sigma})/2$ where $\mathcal{I}$ is
the 2x2 identity matrix and
$\bm{\sigma}=\{\sigma_x,\sigma_y,\sigma_z\}$ are the Pauli
matrices. Self-consistent methods\cite{Kubler,Sandratskii,nord96} for
calculating the electronic structure for non-collinear magnetization
densities have existed for quite some time\cite{Sandratskii2}, and
here we will focus on the specific details for treating non-collinear
magnetization densities within the RS-LMTO-ASA. 
\par
With the recursion method, the local density of states $N(\epsilon)$
where $\epsilon$ is the energy, is obtained as
$N(\epsilon)=-\frac{1}{\pi}\Im\tr\mathcal{G}(\epsilon)$. Here
$\mathcal{G}(\epsilon)$ is the local Green's function
$\mathcal{G}(\epsilon)=(\epsilon-\mathcal{H})^{-1}$, where
$\mathcal{H}$ is the Hamiltonian. Similar to the LDOS, the collinear
magnetic density of states $m(\epsilon)$ can be calculated as
$m(\epsilon)=-\frac{1}{\pi}\Im\tr({\sigma}_z\mathcal{G}(\epsilon))$.  
Since the Pauli spin matrix $\sigma_z$ is diagonal in spin-space, the
collinear magnetic density of states can be calculated using only
diagonal elements of the Green's function. If a generalized
non-collinear magnetization density  
\begin{equation}
\label{eqn1}
\bm{m}(\epsilon)=-\frac{1}{\pi}\Im\tr(\bm{\sigma}\mathcal{G}(\epsilon))
\end{equation}
where $\bm{\sigma}=\{\sigma_x,\sigma_y,\sigma_z\}$ is sought,
evaluation of the off-diagonal parts of the Green's function is in
principle needed. The off-diagonal elements of the Green's function
are possible to obtain by performing the recursion starting from
carefully selected linear combinations of muffin-tin
orbitals\cite{Petrilli90} or perform a computationally more demanding
block recursion calculation\cite{Saha05}. However, in our
implementation we avoid the evaluation of off-diagonal elements by
applying successive unitary transformations $\mathcal{U}$ on the
Hamiltonian, $\mathcal{H}'=\mathcal{UHU}^\dagger$. When the
Hamiltonian is transformed in this way, the Green's function transform
similarly; $\mathcal{G}'=\mathcal{UGU}^\dagger$. 
\par
Using the unitary property $\mathcal{U}^\dagger\mathcal{U}=1$ and the
fact that cyclic permutations of matrix multiplications conserve the
trace of the product, the generalized magnetic density of states
$\bm{m}(\epsilon)$, can be written as 
\begin{equation}
\label{eqn2}
\bm{m}(\epsilon)=-\frac{1}{\pi}\Im tr\{\bm{\sigma}\mathcal{U}^\dagger\mathcal{U}\mathcal{G}\mathcal{U}^\dagger\mathcal{U}\}
=-\frac{1}{\pi}\Im tr\{\bm{\sigma}'\mathcal{G}'\},
\end{equation}
where $\bm{\sigma}'$ is the Pauli matrices after the unitary
transformation. The transformation matrix $\mathcal{U}$ is different
for the three directions, and chosen so that,
$\mathcal{U}\sigma_j\mathcal{U}^\dagger=\sigma'_z$, for $j=x,y,z$, to
yield a diagonal representation. In the trivial case of $j=z$ the
unitary transformation is just the identity matrix. For the other
directions, the unitary transformation corresponds to a spin rotation
where $\mathcal{U}$ can be calculated using spin-$\frac{1}{2}$
rotation matrices. Decomposing the Hamiltonian into a spin-dependent
part, $\mathbf{B}$, and a spin-independent component, $H$, yields that
$\mathcal{U}$ operates only on the spin-dependent part, 
\begin{equation}
\mathcal{H}' = H  + \bm{B} \cdot \mathcal{U}\bm{\sigma}\mathcal{U}^\dagger.
\label{eqn3}
\end{equation}
From the transformed Hamiltonians, $\mathcal{H}'$, the LDOS for the
different directions can then be calculated using the regular
recursion method and the magnetic density along the three directions
can be obtained. From the three orthogonal directions, the local
magnetization axis is calculated and the LDOS for the local spin axis
can be constructed by taking the scalar product of the generalized
magnetic density of states and the local magnetization vector. As all
Hamiltonians are constructed within an {\it ab initio} LMTO-ASA
formalism, all calculations are fully self-consistent, and the spin
densities are treated within the local spin density
approximation\cite{barth72}. Since the recursion procedure is
performed for three orthogonal directions, the computational cost for
each iteration is tripled compared with the collinear implementation
of the RS-LMTO-ASA, but the linear scaling with respect to the number
of nonequivalent atoms is retained. 
\par
The calculations of the transition metal clusters have been performed
by embedding the clusters as a perturbation on a self-consistently
converged perfect Cu(111) surface. The Cu surface has been calculated
using the experimental lattice parameter of Cu.  As is usually the
case for LMTO-ASA methods, the vacuum outside the surface needs to be
simulated by having a number of layers of empty spheres above the Cu
surface in order to provide a basis for the wave-function in the
vacuum and to treat charge transfers correctly. After embedding the
cluster on the surface, the charge and magnetization densities of the
cluster atoms and the neighboring Cu atoms and empty spheres are then
recalculated until self-consistency is obtained while the electronic
structure for atoms far from the cluster are kept unchanged to their
unperturbed values. Structural relaxations have not been included in
this study, so the cluster sites have been placed on the regular fcc
lattice above the Cu surface. Earlier studied on supported transition
metal clusters\cite{Pick2003} have shown that structural relaxations
can change the magnetic properties of the clusters. On the other hand,
in an experimental situation, small clusters as those considered in
this study are usually constructed in an out-of-equilibrium situation
by manipulation with an STM tip and calculated equilibrium geometries
might therefore not be relevant. The most relevant relaxation for
these kind of artificially created clusters would be the distance
between the cluster atoms and the substrate atoms and since a noble
metal substrate is used in this study, the interaction between
clusters and substrate play a lesser role compared to the interactions
between the cluster atoms. The clean Cu(111) surface has been modeled
by a large ($>$5000) slab of atoms and the continued fraction, that
occurs in the recursion method, have been terminated with the
Beer-Pettifor\cite{Beer84} terminator after 30 recursion levels.  
\par
The non-collinear calculations have been performed without including
the spin-orbit coupling. Since this term is neglected, a preferred
spin axis does not exist in the system and the magnetic structures are
thus only converged with respect to the directions of the magnetic
moments relative to the other spin moments in the cluster. In order to
minimize the risk of finding magnetic orderings that correspond to
only a local minimum, several starting guesses where used for each system.
\par
The calculated magnetic structures can be analyzed in terms of the
exchange interactions $J_{ij}$ between spins on atoms situated
at sites $i$ and site $j$. A well known connection between the
exchange interactions and the magnetic ordering is given by the
classical Heisenberg Hamiltonian,
\begin{equation}
\mathcal{H_{H}} = -\sum_{i,j,i\not=j}J_{ij}cos{\theta_{ij}} ,
\label{eqn4}
\end{equation}
where $\theta_{ij}$ is the angle between the magnetic moment on site
$i$ and site $j$. Note that in Eqn.~\ref{eqn4}, the magnitude of the
spins have been incorporated into the effective $J_{ij}$
interactions. 
\par
In this work, we have calculated exchange interactions directly
using the Liechtenstein formula\cite{Liechtenstein87} as
implemented in the RS-LMTO-ASA\cite{Frota-Pessoa00} for a large selection of the
considered clusters. The $J_{ij}$'s shown in this study, have been
obtained from the ferromagnetic configuration of the clusters. Other magnetic configurations typically result in different values of the $J_{ij}$'s, although the signs are seldom changed.\cite{soniaangela}
If the exchange interactions would be independent of $\theta_{ij}$,
the magnetic structure could in principle be calculated by minimizing
the Heisenberg Hamiltonian with $J_{ij}$'s calculated from a ferromagnetic configuration. This is not the case for the systems
considered in this work, which motivates a full non-collinear calculation of the magnetic structures. However, on a qualitative level the cause of the magnetic ordering, e.g. the effect of frustration or the competition between nearest and next-nearest interactions, can still be discussed in
terms of the calculated exchange interactions.
\par
For a selection of Cr clusters, which are discussed in
Sec.~\ref{crclusters}, our calculated magnetic structures has been
compared to the structure found by minimizing the Heisenberg
Hamiltonian for a fixed configuration of $J_{ij}$'s, where only
nearest-neighbour interactions are finite. The minimization of the
Heisenberg Hamiltonian for these clusters has been performed by a
genetic search algorithm\cite{genprice}.  
\section{Results}
\subsection{Fe clusters}
\begin{figure}
\begin{center}
\subfigure[]{
\includegraphics*[width=0.15\textwidth]{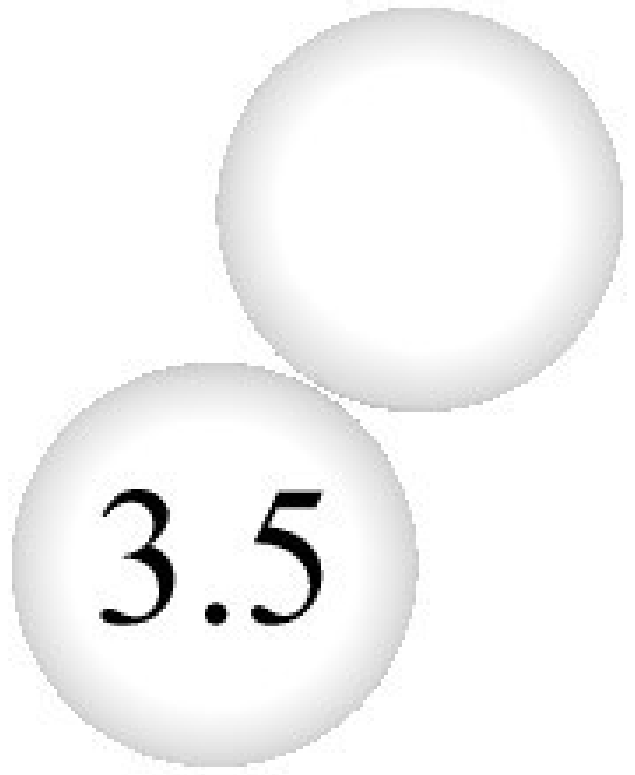}
\label{figfe1:a}}
\subfigure[]{
\includegraphics*[width=0.15\textwidth]{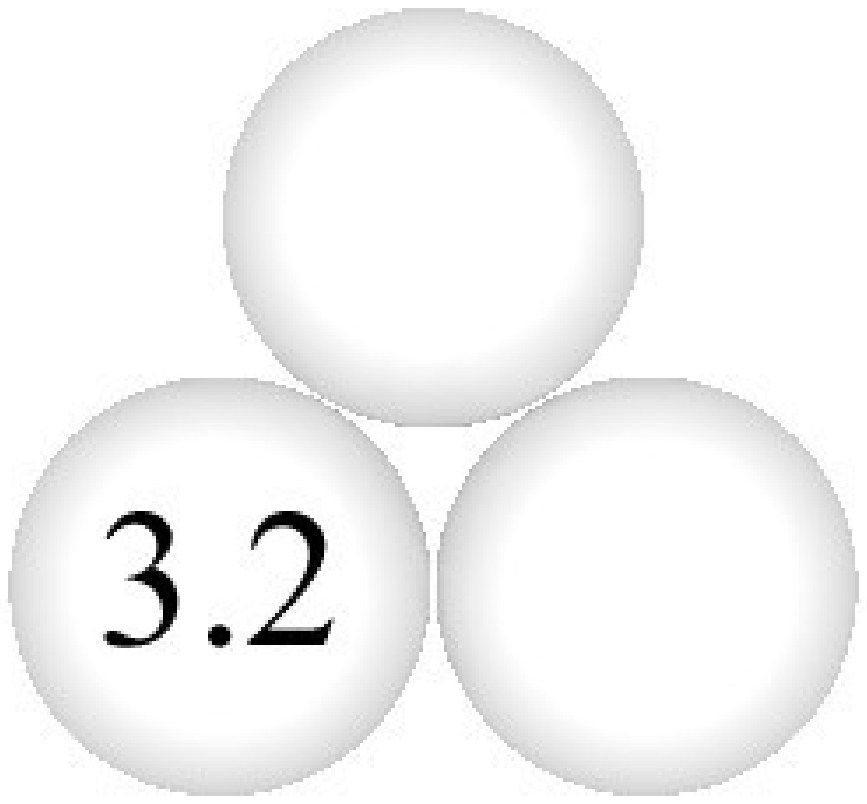}
\label{figfe1:b}}\\
\subfigure[]{
\includegraphics[width=0.15\textwidth]{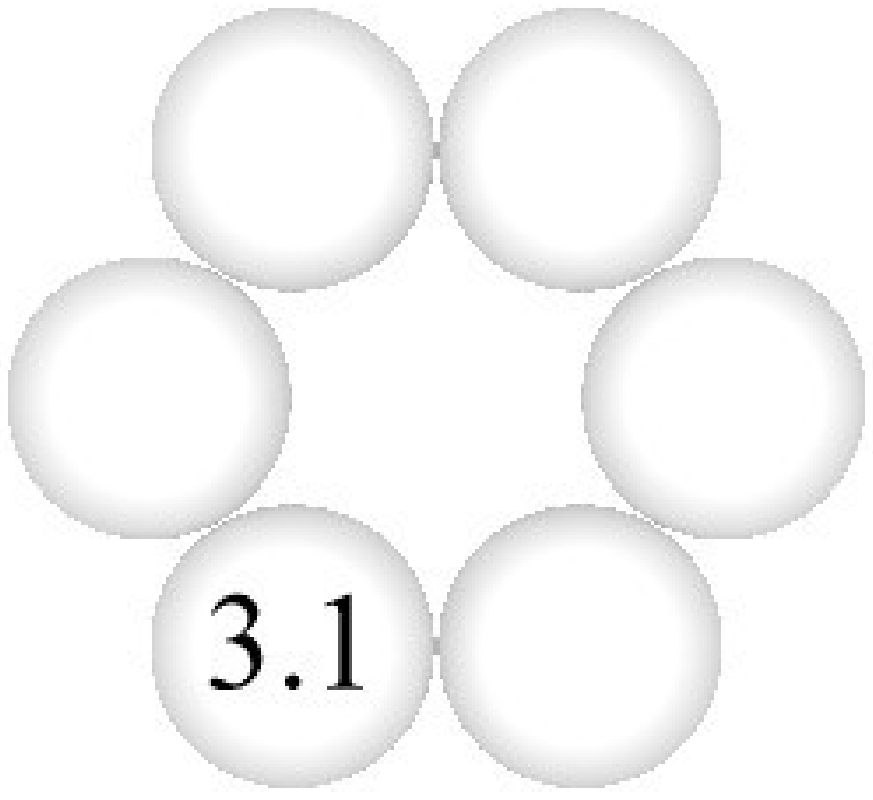}
\label{figfe1:c}}
\subfigure[]{
\includegraphics*[width=0.15\textwidth]{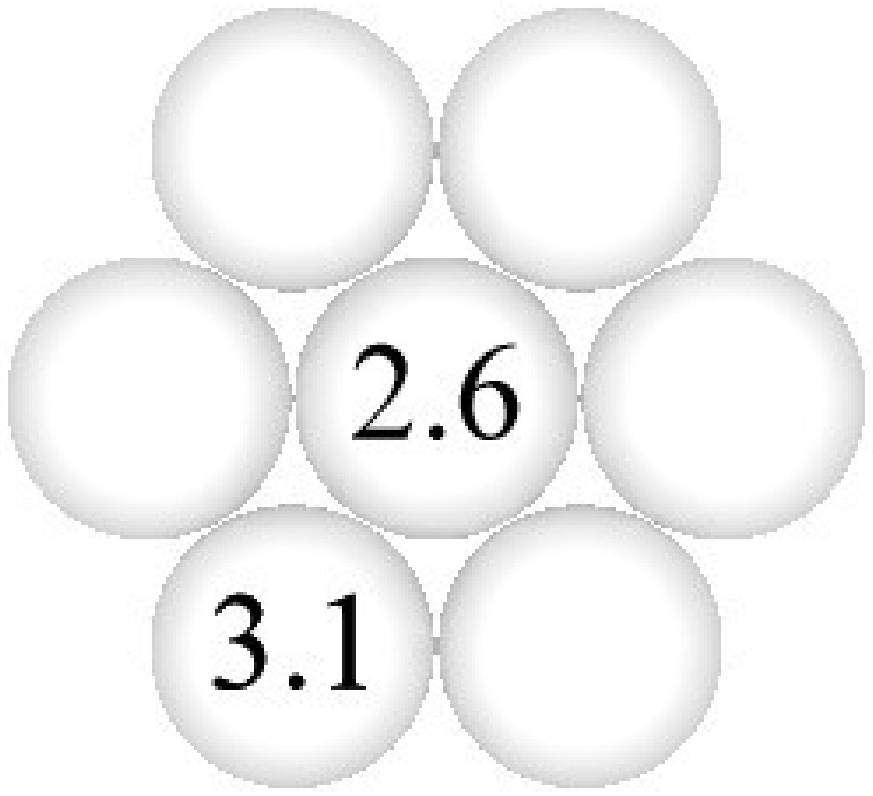}
\label{figfe1:d}}\\
\subfigure[]{
\includegraphics*[width=0.15\textwidth]{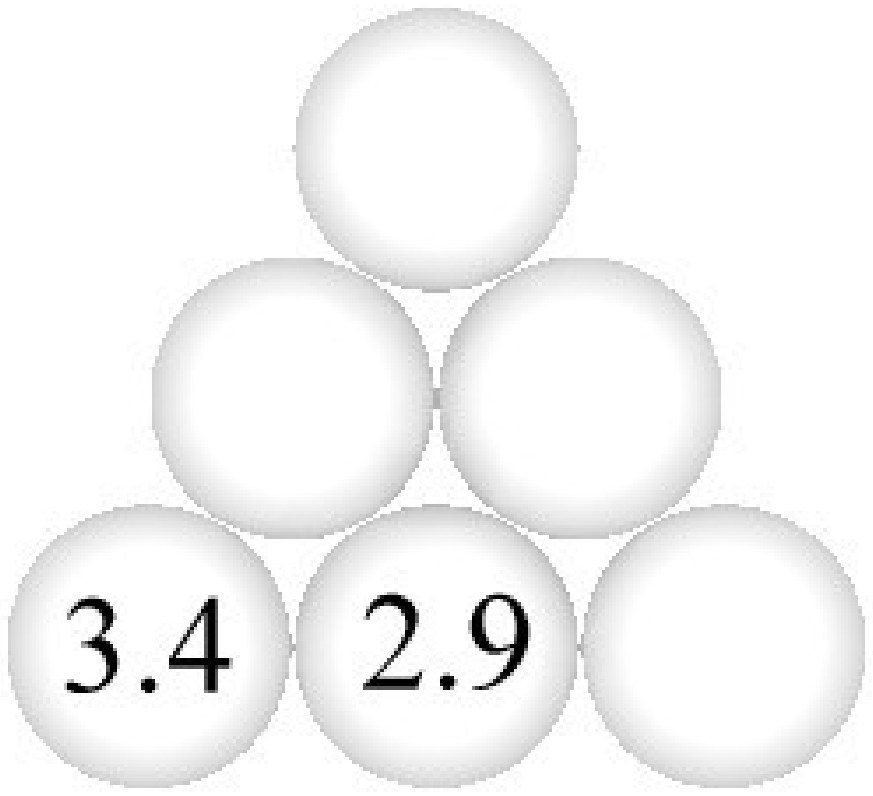}
\label{figfe1:e}}
\subfigure[]{
\includegraphics*[width=0.15\textwidth]{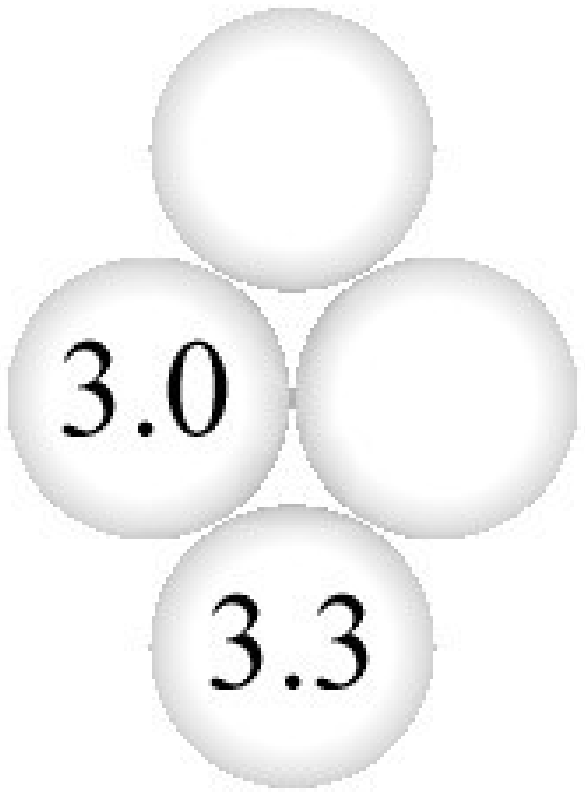}
\label{figfe1:f}}\\
\caption{\label{figfe1} The geometries for clusters of Fe atoms on a
Cu(111) surface. All Fe clusters are found to exhibit a ferromagnetic
ground state regardless of the cluster structure. The numbers indicate
the atom projected spin moment of the different atoms.} 
\end{center}
\end{figure}

In Fig.~\ref{figfe1} the magnetic structure of several Fe clusters are
shown. Regardless of the geometry of the studied Fe clusters, we find the
magnetic ordering in the clusters to always be ferromagnetic. The
collinear magnetic structure for these clusters, which can be put in
contrast with the non-collinear ordering found for fcc structured Fe
clusters embedded in bulk Cu\cite{Tsunoda89}, could be caused by the
fact that the decreased coordination of the surface clusters lead to a
high-spin state which favours ferromagnetic coupling between
neighbouring Fe atoms\cite{Ujfalussy96}. This is also consistent with
an analysis by Lizarraga {\it et al.}\cite{Lizarraga04}. It should be
pointed out here that large magnetic moments do not automatically lead
to collinear magnetism. As we will see in the section below, Mn is an
example where large moments result in an antiferromagnetic interatomic
exchange coupling, which on a frustrated geometry lead to
non-collinear magnetism. In the case of Fe the large calculations
result in large moments and a ferromagnetic interatomic exchange
coupling.  
\par
The spin moments for the Fe atoms in the clusters shown in
Fig.~\ref{figfe1} range between 3.45 $\mu_B$ for the atoms in the dimer
to 2.56 $\mu_B$ for the central atom in the seven atom cluster
displayed in Fig.~\ref{figfe1:d}. It has been shown for Co clusters
in Cu(001)\cite{Klautau05} and Fe clusters on Ni and Cu
surfaces\cite{Mavropoulos06}, that the magnetic moment has a linear
behavior as function of the number of cluster neighbors around the
site. The spin moments of the Fe clusters on Cu(111) of
Fig.~\ref{figfe1} show a similar trend and depend almost linearly on
the number of nearest Fe neighbours. For these ferromagnetic Fe
clusters, the orbital moments were calculated and they where also found to 
depend on the number of nearest neighbours. The largest orbital moment was
found to be 0.15 $\mu_B$ per atom for the atoms in the dimer, and the
smallest orbital moment is 0.06 $\mu_B$ for the central atom in the
cluster shown in Fig.~\ref{figfe1:d}.

Due to the strong correlation between the magnetic moment for the
atoms in the Fe clusters and the number of Fe neighbours, the magnetic
moments obtained above can in principle be used
to predict the total magnetic moment of any Fe cluster as long as the
shape is determined and the cluster is planar. This would indicate
that the magnetic moment per atom for a perfect monolayer of Fe atoms
on a Cu(111) surface would be $\sim$~2.7 $\mu_B$ which is in good
agreement with earlier calculations of Fe monolayers on
Cu\cite{Kruger00,Spisak03,Hjortstam96,Fernando88}. Our findings of
ferromagnetic coupling indicate that a single Fe monolayer would be
ferromagnetic in agreement with Ref.~\onlinecite{Kruger00} whereas
Ref.~\onlinecite{Spisak03} found that a single row antiferromagnetic
order would be the most stable magnetic configuration. This
discrepancy may be explained by the use of different lattice
parameters in Refs.~\onlinecite{Kruger00} and \onlinecite{Spisak03}. 
\subsection{Mn clusters}
\begin{figure}
\begin{center}
\subfigure[]{
\includegraphics*[width=0.18\textwidth]{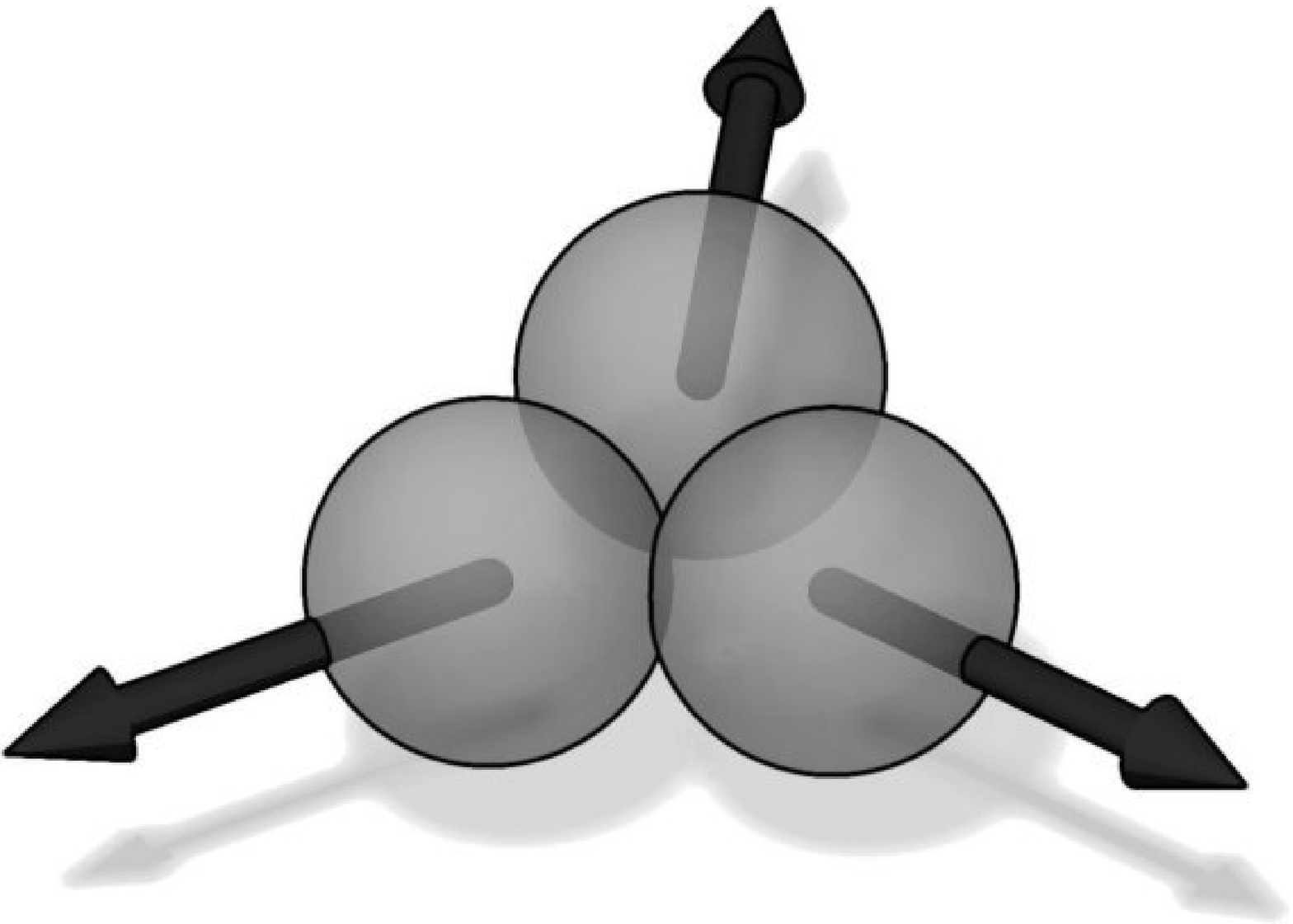}
\label{figmnt1:a}}
\subfigure[]{
\includegraphics*[width=0.15\textwidth]{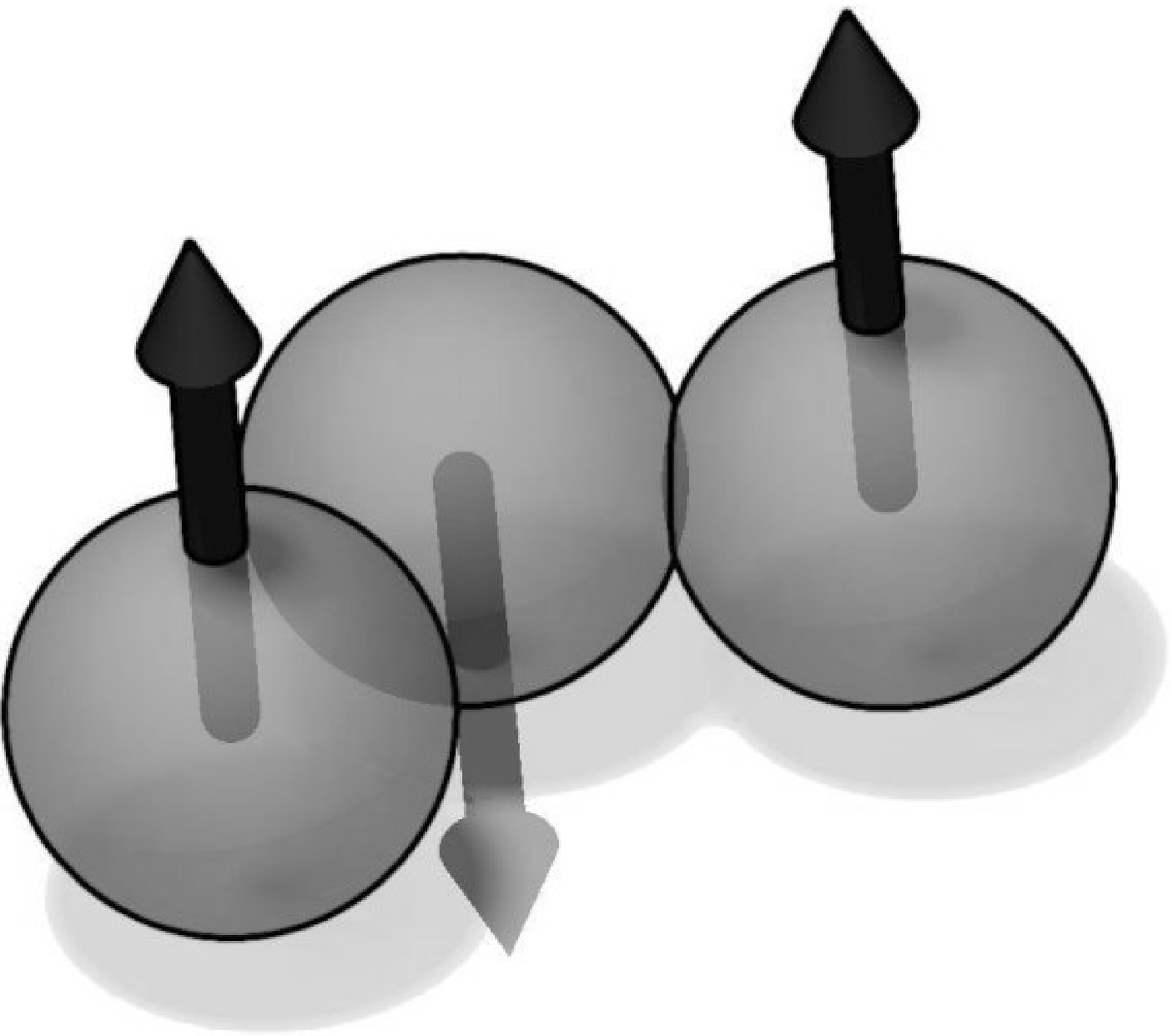}
\label{figmnt1:b}}\\
\subfigure[]{
\includegraphics[width=0.15\textwidth]{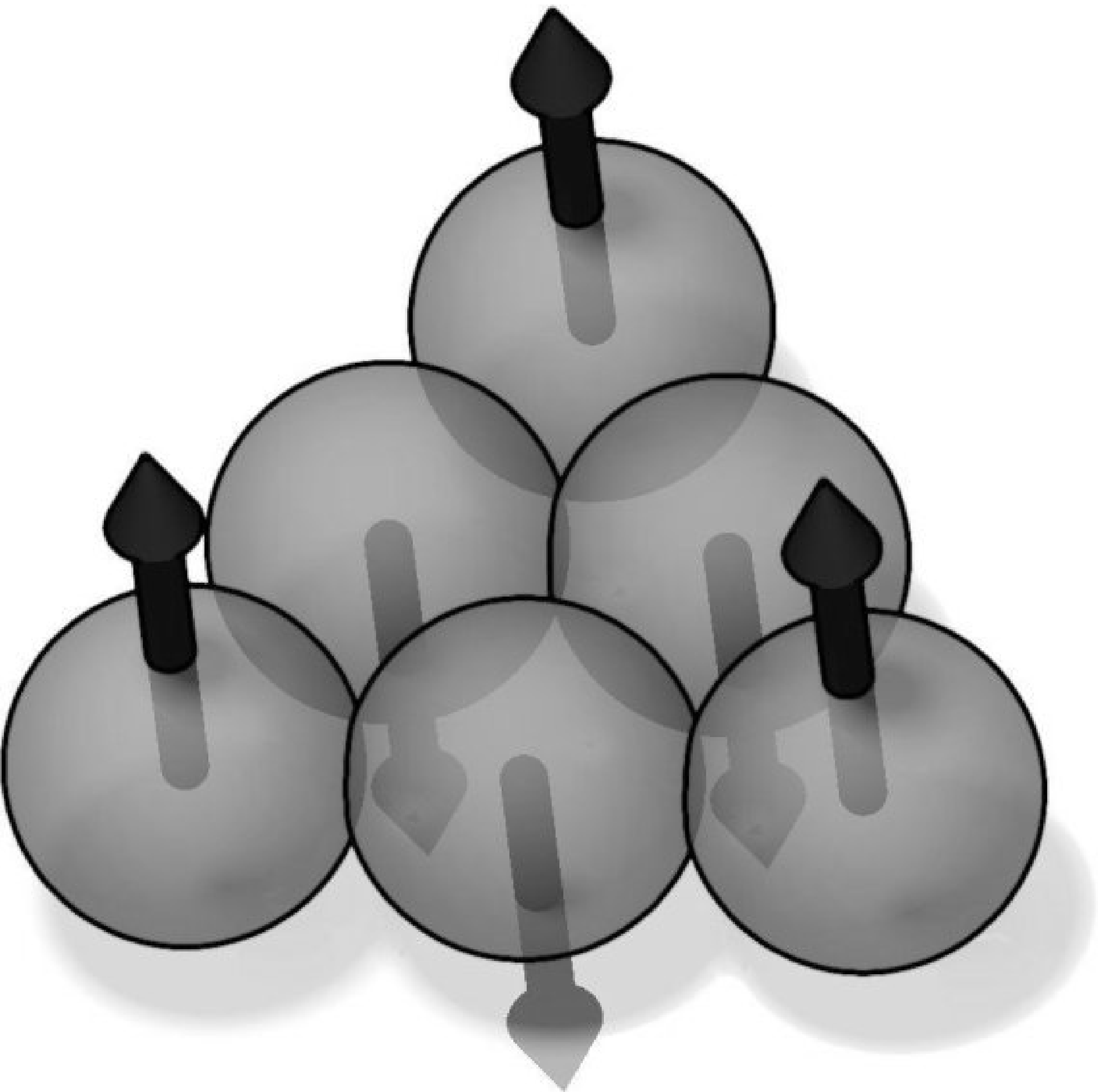}
\label{figmnt1:c}}
\subfigure[]{
\includegraphics*[width=0.21\textwidth]{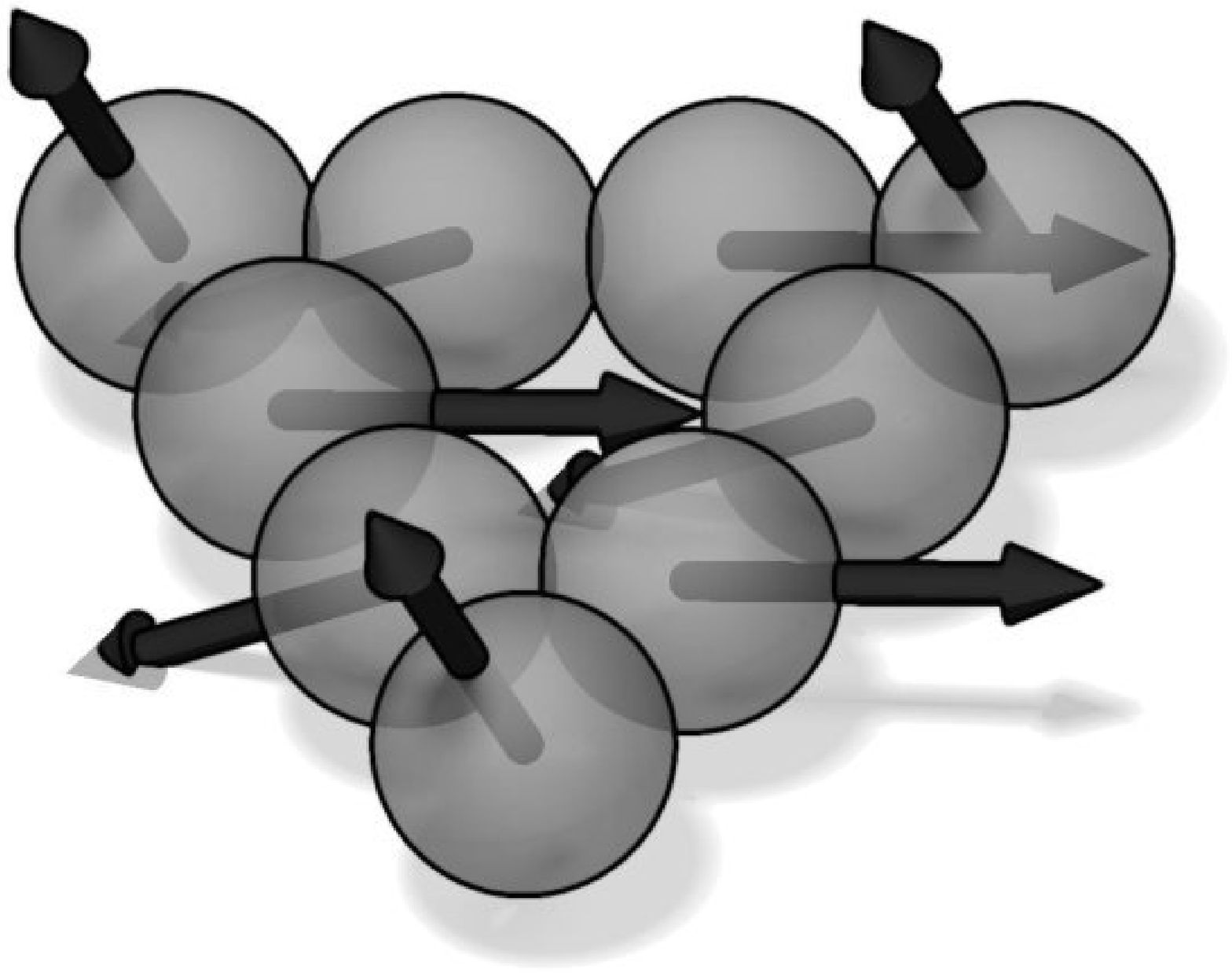}
\label{figmnt1:d}}\\
\subfigure[]{
\includegraphics*[width=0.23\textwidth]{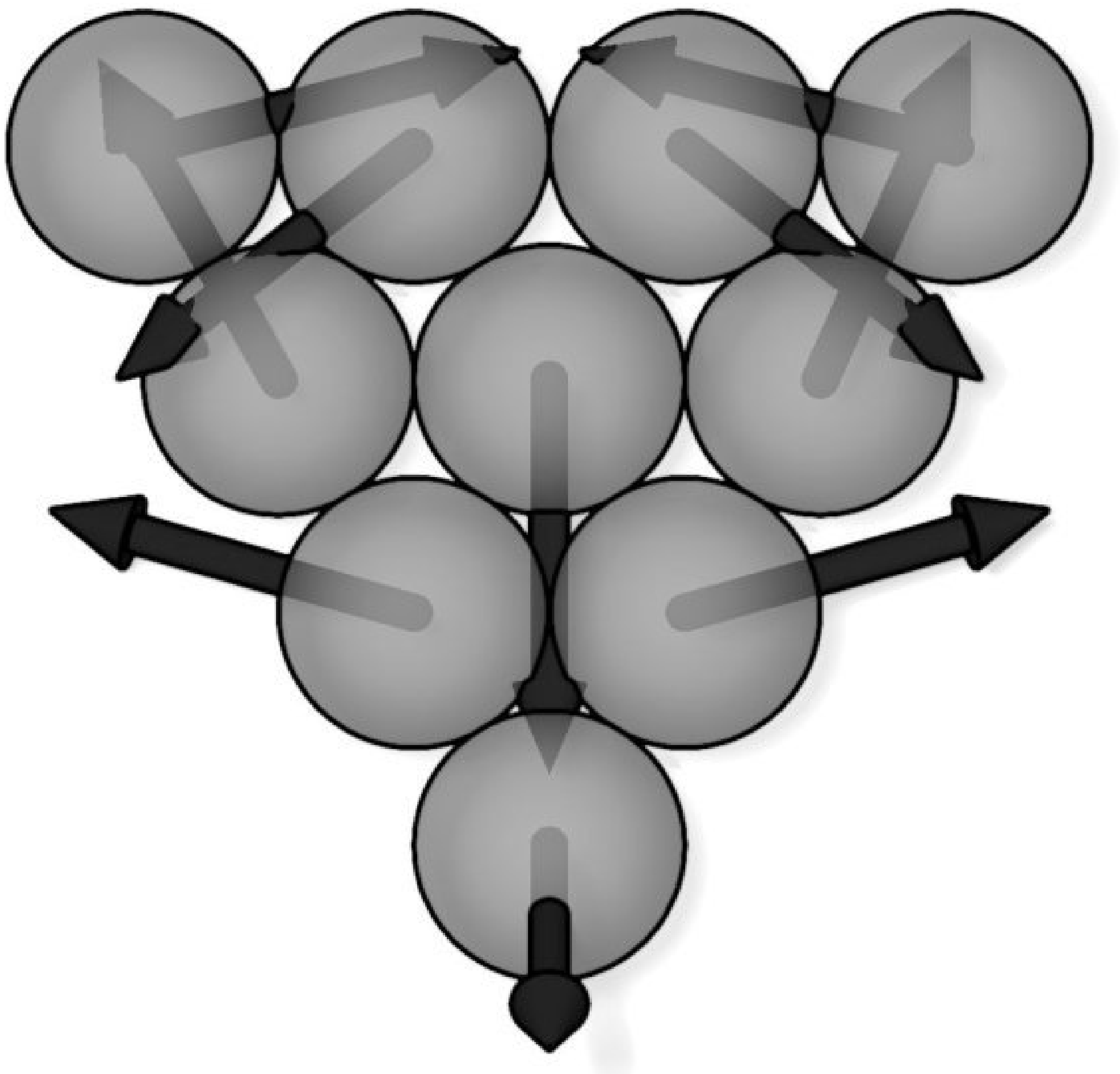}
\label{figmnt1:e}}
\caption{\label{figmnt1} The calculated magnetic ordering for
triangular Mn clusters on a Cu(111) surface. For all geometries except
the isosceles triangle shown in (b) and the six atom triangle in (c),
non-collinear solutions are obtained due to magnetic frustration. } 
\end{center}
\end{figure}

In a previous paper\cite{Bergman06} we showed that due to
antiferromagnetic coupling between nearest neighbour atoms in Mn
clusters deposited on Cu one finds either a collinear
antiferromagnetic structure or, if frustration occurs due to the
cluster geometry, a non-collinear magnetic structure. A collection of
frustrated cluster geometries with triangular shapes is shown in
Fig.~\ref{figmnt1}. The magnetic moments obtained for each atom of the
studied Mn cluster are shown in Table.~\ref{table:mn1}. For the equilateral triangle in
Fig.~\ref{figmnt1:a} a non-collinear arrangement with angle of
120$^\circ$ between the magnetic moments is the most stable
configuration. The total energy difference between the stable
non-collinear solution and a frustrated collinear antiferromagnetic solution,
with two spins parallel to each other and anti-parallel to the third spin
is 13 meV/atom. The ferromagnetic solution was found to have an
energy of 102 meV/atom higher than the stable non-collinear
solution. The isosceles triangle shown in Fig.~\ref{figmnt1:b} has an
antiferromagnetic collinear ground state which indicates that the
exchange coupling between the two Mn atoms furthest from each other is
either very small compared to the antiferromagnetic nearest-neighbour
exchange coupling or has the opposite sign (i.e. ferromagnetic). 
As the size of the triangular clusters increases (Figs.~\ref{figmnt1:c} to
\ref{figmnt1:e}), the behaviour becomes more intricate. Although the
six-atom triangle (Fig.~\ref{figmnt1:c}) by the analogous geometry as
the three atom triangle in Fig.~\ref{figmnt1:a}, could be expected to
align in a structure with 120$^\circ$ between neighbouring atoms, it
is in fact a collinear antiferromagnetic order that is the most stable
solution. The energy difference between the collinear structure and a
non-collinear structure was found to be 18 meV/atom. The cause for the
preferred collinear order is the different environment for the corner
atoms who only have two nearest neighbours compared to the three
central atoms who have four nearest neighbours each. The reduced
coordination for the corner atoms causes their antiferromagnetic
exchange coupling to nearest neighbours to be enhanced. The calculated
J$_{ij}$s confirm this behaviour since the strength of the exchange
interaction between a corner atom and a nearest-neighbour amounts to
-27 meV while it is -12 meV between two central atoms.
 A similar mechanism can be expected for the nine-atom cluster displayed
in Fig.~\ref{figmnt1:d}, but for this geometry the atoms that are not
situated at the corners of the triangle have three nearest neighbours
due to the hole in the middle of the cluster. Therefore the difference
in the local geometry is smaller between the corner atoms and the
central atoms which leads to a more delicate balance of the exchange
couplings. The resulting structure has the moments pointing in three
different directions instead of two directions which would be the case for
a collinear antiferromagnetic solution. The angle between two neighbouring
central atoms is 152$^\circ$ while the angle between a
corner atom and a nearest-neighbour is 104$^\circ$. Contrary to what
was found for the six atom cluster in Fig.~\ref{figmnt1:c}, it thus
appears that the exchange coupling is larger between central atoms
then between a corner atom and a central atom. This behaviour is
supported by the calculated J$_{ij}$s where the coupling between an
corner atom and a nearest-neighbour is -9meV while the exchange
coupling is found to be -36 meV between two central atoms. The
non-collinear solution for the nine atom cluster has a total energy
which is 13me V/atom lower than a collinear antiferromagnetic solution.
\par
The analysis of the final ten-atom triangle shown in
Fig.~\ref{figmnt1:e} is even more complicated. From a geometrical
view, this cluster has three nonequivalent sites; the three corner
atoms, the six atoms neighbouring to the corner atoms and the central
atom. The magnetic structure does however have a lower symmetry that
can be described by decomposing the cluster into the central atom and
three 'sub-triangles', consisting of the three atoms closest to
each corner of the cluster. Within each sub-triangle, the
three atoms couple to each other in a geometry that 
resembles the 120$^\circ$ structure of a single three atom triangular
cluster, but since exchange interactions from other neighbouring atoms
are present as well, the angles between the moments in these
sub-triangles vary between 146$^\circ$ and 104$^\circ$.
All angles between the moments for the atoms in the ten atom
cluster can be seen in Table~\ref{table:mn1e}.
The influence of the different number of neighbours for the cluster
atoms determine their magnetic moments where the corner
atoms have a magnetic moment of 4.3 $\uB$, the central atom has
2.6 $\uB$ and the magnetic moment for the remaining six atoms is
3.6 $\uB$. 
\begin{table}
\caption{\label{table:mn1} Magnetic moments (in $\mu_B$) for atoms in
the clusters displayed in Fig.~\ref{figmnt1:a}-\ref{figmnt1:e}. The atoms are numbered
in the left column, starting from the  leftmost atom, increasing around the cluster in the
clockwise direction and, for the largest cluster, ending with the central atom. }
\begin{center}
\begin{ruledtabular}
\begin{tabular}{cccccc}
 ~ & 2(a) & 2(b) & 2(c) & 2(d) & 2(e) \\
\hline
1 & 4.25& 4.54&  4.33&  4.15&  4.25\\
2 & 4.25& 4.23&  3.57&  3.77&  3.56\\
3 & 4.25& 4.54&  4.33&  3.77&  3.56\\
4 &  -  &  -  &  3.57&  4.15&  4.25\\
5 &  -  &  -  &  4.37&  3.77&  3.56\\
6 &  -  &  -  &  3.57&  3.77&  3.56\\
7 &  -  &  -  &   -  &  4.15&  4.25\\
8 &  -  &  -  &   -  &  3.77&  3.56\\
9 &  -  &  -  &   -  &  3.77&  3.56\\
10&  -  &  -  &   -  &   -  &  2.64\\
\end{tabular}
\end{ruledtabular}
\end{center}
\end{table}
\begin{table}
\caption{\label{table:mn1e} The magnetic configuration described by angles
between moments for atoms in the cluster displayed in
Fig.~\ref{figmnt1:e}. The atoms are numbered starting from the
leftmost atom, increasing around the cluster in the clockwise
direction and ending with the central atom. }
\begin{center}
\begin{ruledtabular}
\begin{tabular}{ccccccccccc}
 Atom & 1 & 2 & 3 & 4 & 5 & 6 & 7 & 8 & 9 & 10 \\%& Moment\\
\hline
1 &  0&  146&   52&  147&   57&   20&   90&  150&  104&  103\\
2 &146&    0&   98&   52&  155&  158&   56&   60&  107&   52\\
3 & 52&   98&    0&  146&  107&   60&   56&  158&  155&   52\\
4 &147&   52&  146&    0&  104&  150&   90&   20&   57&  103\\
5 & 57&  155&  107&  104&    0&   47&  138&   96&   49&  150\\
6 & 20&  158&   60&  150&   47&    0&  108&  141&   96&  107\\
7 & 90&   56&   56&   90&  138&  108&    0&  108&  138&   53\\
8 &150&   60&  158&   20&   96&  141&  108&    0&   47&  107\\
9 &104&  107&  155&   57&   49&   96&  138&   47&    0&  150\\
10&103&   52&   52&  103&  150&  107&   53&  107&  150&    0\\
%1 &     0&  104&   57&   96&   49&  107&  138&  155&   47&  150\\%&4.25\\
%2 &   104&    0&  147&   20&   57&  146&   90&   52&  150&  103\\%&3.54\\
%3 &    57&  147&    0&  150&  104&   52&   90&  146&   20&  103\\%&3.54\\
%4 &    96&   20&  150&    0&   47&  158&  108&   60&  141&  107\\%&4.25\\
%5 &    49&   57&  104&   47&    0&  155&  138&  107&   96&  150\\%&3.54\\
%6 &   107&  146&   52&  158&  155&    0&   56&   98&   60&   52\\%&3.54\\
%7 &   138&   90&   90&  108&  138&   56&    0&   56&  108&   53\\%&4.25\\
%8 &   155&   52&  146&   60&  107&   98&   56&    0&  158&   52\\%&3.54\\
%9 &    47&  150&   20&  141&   96&   60&  108&  158&    0&  107\\%&3.54\\
%10&   150&  103&  103&  107&  150&   52&   53&   52&  107&    0\\%&2.64\\
\end{tabular}
\end{ruledtabular}
\end{center}
\end{table}
\par
\begin{figure}
\begin{center}
\includegraphics*[width=0.45\textwidth]{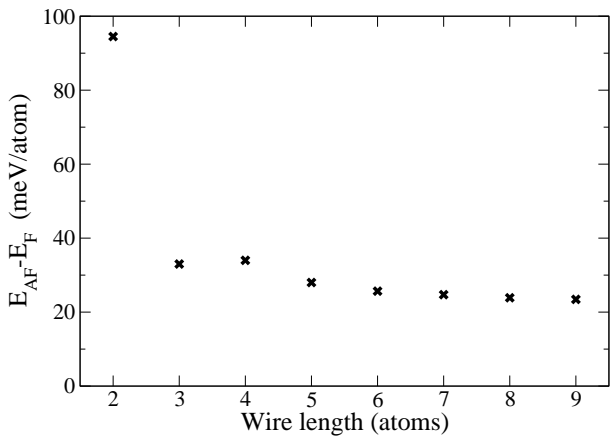}
\caption{\label{fig:ewire} Total energy differences between
antiferromagnetic, $E^{AF}$, and ferromagnetic, $E^{FM}$,
configurations of wires of Mn atoms, oriented along a $(1\bar{1}0)$
direction on a Cu(111) surface.} 
\end{center}
\end{figure}

Atomic wires constitute a group of nanostructures that has attracted a
lot of attention\cite{Gambardella2001,Ujfalussy04,Lee04}. We have
calculated the magnetic structure for wires of Mn atoms with lengths
between two to nine atoms. The wires are oriented along a $1\bar{1}0$
direction on the Cu surface. 
The total energy differences, per cluster atom, between the
antiferromagnetic and the ferromagnetic magnetic configurations for
the Mn wires are shown in Fig.~\ref{fig:ewire}. A large energy
difference of 96 meV/atom is found for the dimer whereas the energy
differences for the longer wires are significantly smaller. If only
nearest neighbour interactions played role one would expect an energy
difference (per atom) between ferromagnetic and antiferromagnetic
coupling with a functional form J(1-$\frac{1}{N}$), where N is the
number of atoms in the wire. Hence for long chains this energy
difference should be equal to J whereas the energy difference would
continuously become smaller and reach the value J/2 for N=2. The data
in Fig.~\ref{fig:ewire} does not display this trend, which
suggests that next-nearest neighbour interactions are important and/or
that the value of J depends on the number of atoms of the cluster. 
\par
\par
For the dimer and trimer only a collinear
antiferromagnetic solution is found, whereas, for longer chains a
slightly canted non-collinear order is also found. For almost all of
the longer chains, the non-collinear solutions are unstable, but
resembles the collinear antiferromagnetic solution closely, both in energy, in
all cases less than 0.5 meV/atom, and in the angular difference, where
the deviation from the collinear structure is smaller than 3$^\circ$
per atomic pair.
\par
\begin{figure}
\begin{center}
\includegraphics*[width=0.2\textwidth]{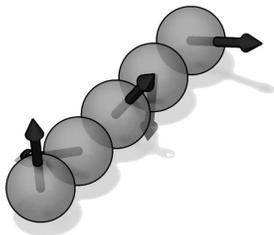}
\caption{\label{figmnm1} Calculated magnetic configurations for a five
atomic Mn wire, oriented along a $(1\bar{1}0)$ direction on a Cu(111)
surface.} 
\end{center}
\end{figure}

The Mn pentamer is however an interesting exception from the behavior of the
other wires, and for this system, the ground state is actually found to
be a non-collinear configuration, shown in Fig.~\ref{figmnm1}.  The
non-collinear configuration for the pentamer 
can be described by the angle between an edge atom and its nearest
neighbour, which is 170$^\circ$, and the angle between two
neighbouring central atoms, which is 155$^\circ$. The energy
difference between the non-collinear and the antiferromagnetic
solutions for the pentamer is 2 meV/atom. 
\par
The straight wires does not have geometries that cause
frustration in the same way as the triangular clusters mentioned
earlier, so a probable cause for the non-collinearity of the Mn pentamer
is the competition between ferromagnetic and antiferromagnetic exchange interactions
between the different atoms in the cluster. Since the
nearest-neighbour interactions are always antiferromagnetic for the Mn
clusters (at least for the nearest-neighbour distance used in this
study), more long-range interactions must play a role in destabilizing
the collinear magnetic state. 
In order to examine the size and range of the exchange interactions,
we have calculated exchange coupling parameters $J_{ij}$ for the five
atom wire shown in Fig.~\ref{figmnm1}. The values of the exchange
parameters are shown in Table~\ref{tab:table1} where the $i$ and $j$
are chosen so that site 1 and 5 are the edge atoms and site 3 is the
central atom. Hence site 2 is nearest neighbour to site 1 and 3 and
the nearest neighbours for site 4 is site 3 and 5. 
\begin{table}
\caption{\label{tab:table1}Calculated exchange parameters $J_{ij}$ (in
meV) for the Mn pentamer shown in Fig.~\ref{figmnm1}. The atoms are
numbered from left to right.} 
\begin{ruledtabular}
\begin{tabular}{cccccc}
 $i \backslash j$ & 1 & 2 & 3 & 4 & 5\\
\hline
  1 & - & -34 & -5.0 & 6.3 & -3.6\\
  2 & -34 & - & -12 & -11 & 6.3\\
  3 & -5.0 & -12 & - & -12 & -5.0\\
  4 & 6.3 & -11 & -12 & - & -34\\
  5 & -3.6 & 6.3 & -5.0 & -34 & -\\
\end{tabular}
\end{ruledtabular}
\end{table}
It may be observed that the exchange interactions are strongest and
antiferromagnetic between nearest neighbours, with a smaller long
range interaction that oscillates between ferromagnetic and
antiferromagnetic coupling. The largest magnitude for the exchange
interaction is obtained for $J_{12}$ and $J_{45}$, i.e. between an
edge atom and its nearest neighbour. Furthermore,
Table~\ref{tab:table1} shows that although the nearest neighbour
interactions have the largest magnitude, the more long ranged
interactions always seem to counteract the nearest neighbour
interactions. This might not be obvious from the values in
Table~\ref{tab:table1} but as a clarifying example we can examine the
exchange interactions between atom 1 and the other atoms. The negative
nearest neighbour interactions in the pentamer would prefer an
antiferromagnetic order so that atom 1 would be ferromagnetically
coupled to atom 3 and atom 5, and antiferromagnetically coupled to
atoms 2 and 4. However, the calculated exchange interactions in
Table~\ref{tab:table1} show that the $J_{13}$ and $J_{15}$ are in fact
negative while $J_{14}$ is positive, thus competing against the
antiferromagnetic nearest-neighbour ordering, which results in the
non-collinear magnetic state shown in Fig.~\ref{figmnm1}. 
\subsection{Cr clusters\label{crclusters}}
\begin{figure}
\begin{center}
\subfigure[]{
\includegraphics*[width=0.12\textwidth]{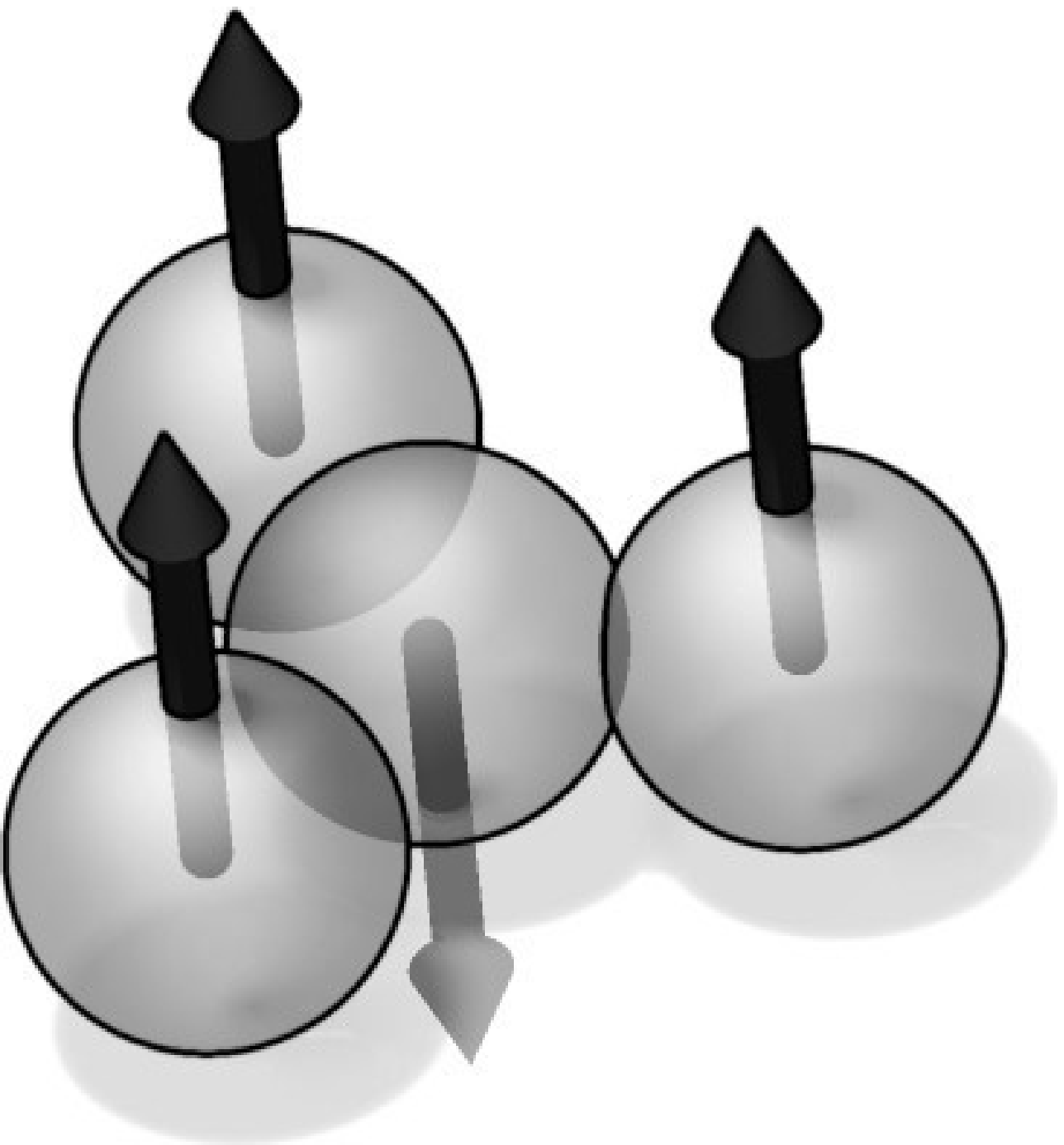}
\label{figcr1:a}}
\subfigure[]{
\includegraphics*[width=0.12\textwidth]{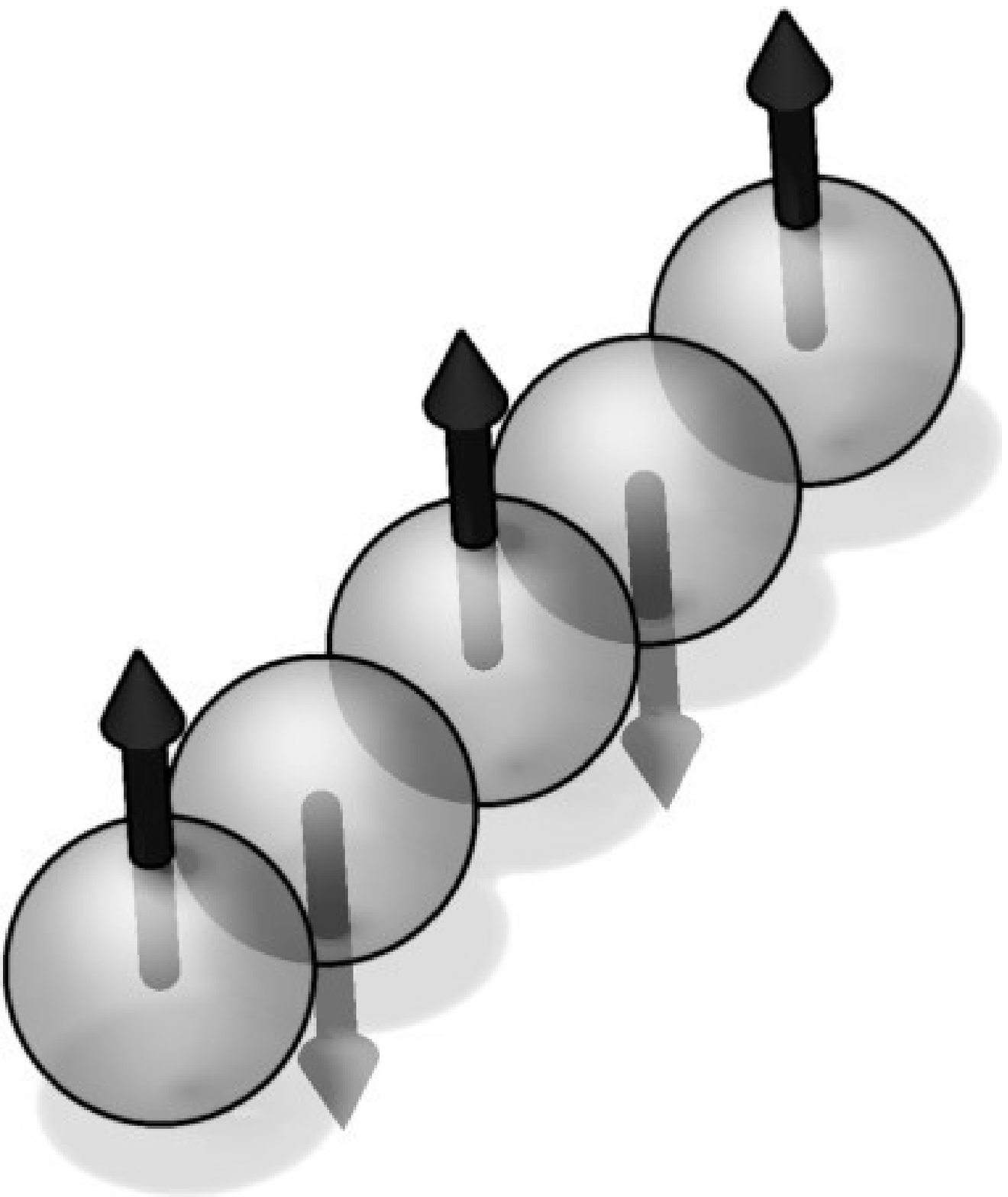}
\label{figcr1:b}}\\
\subfigure[]{
\includegraphics*[width=0.15\textwidth]{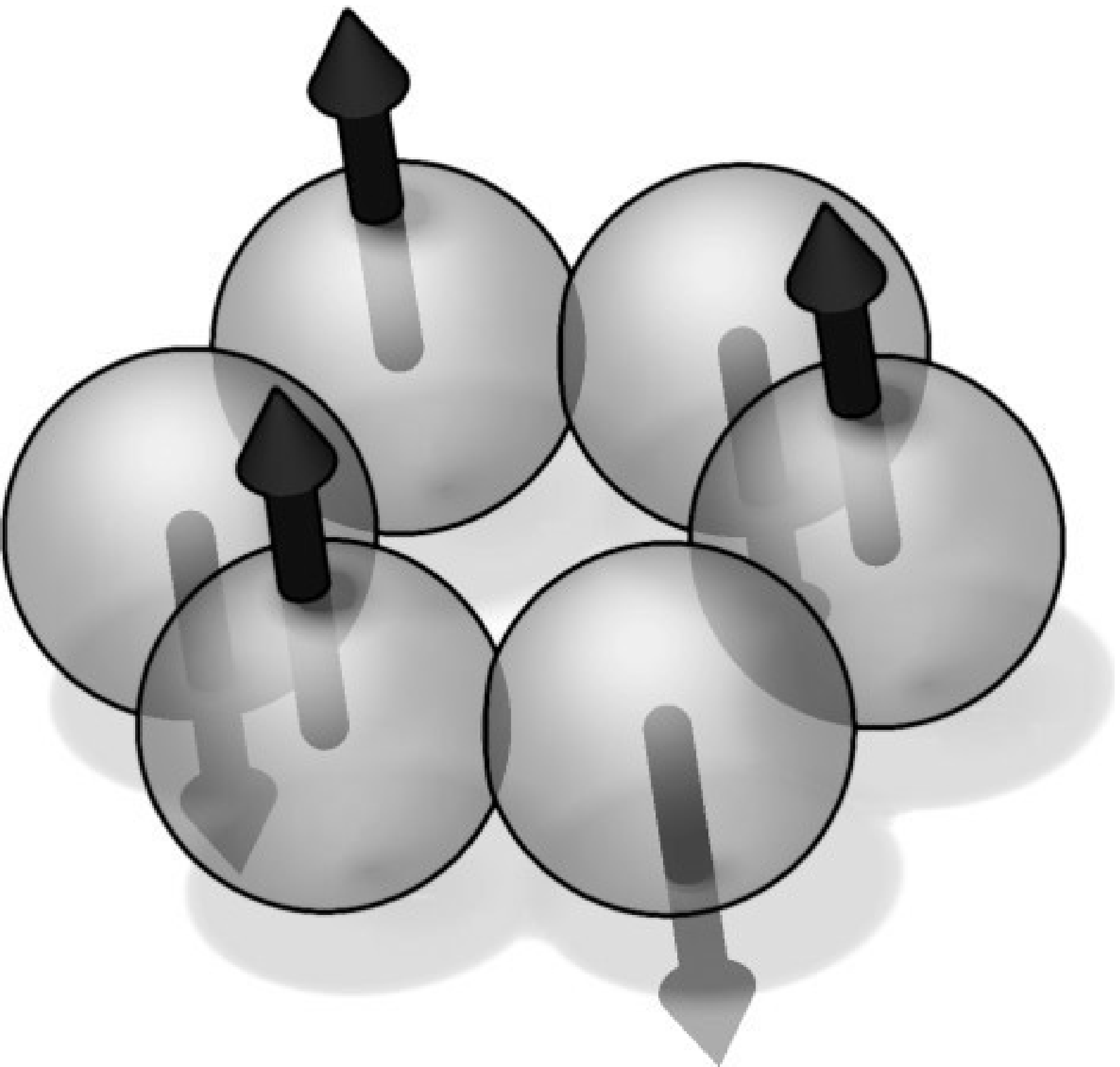}
\label{figcr1:c}}
\subfigure[]{
\includegraphics*[width=0.15\textwidth]{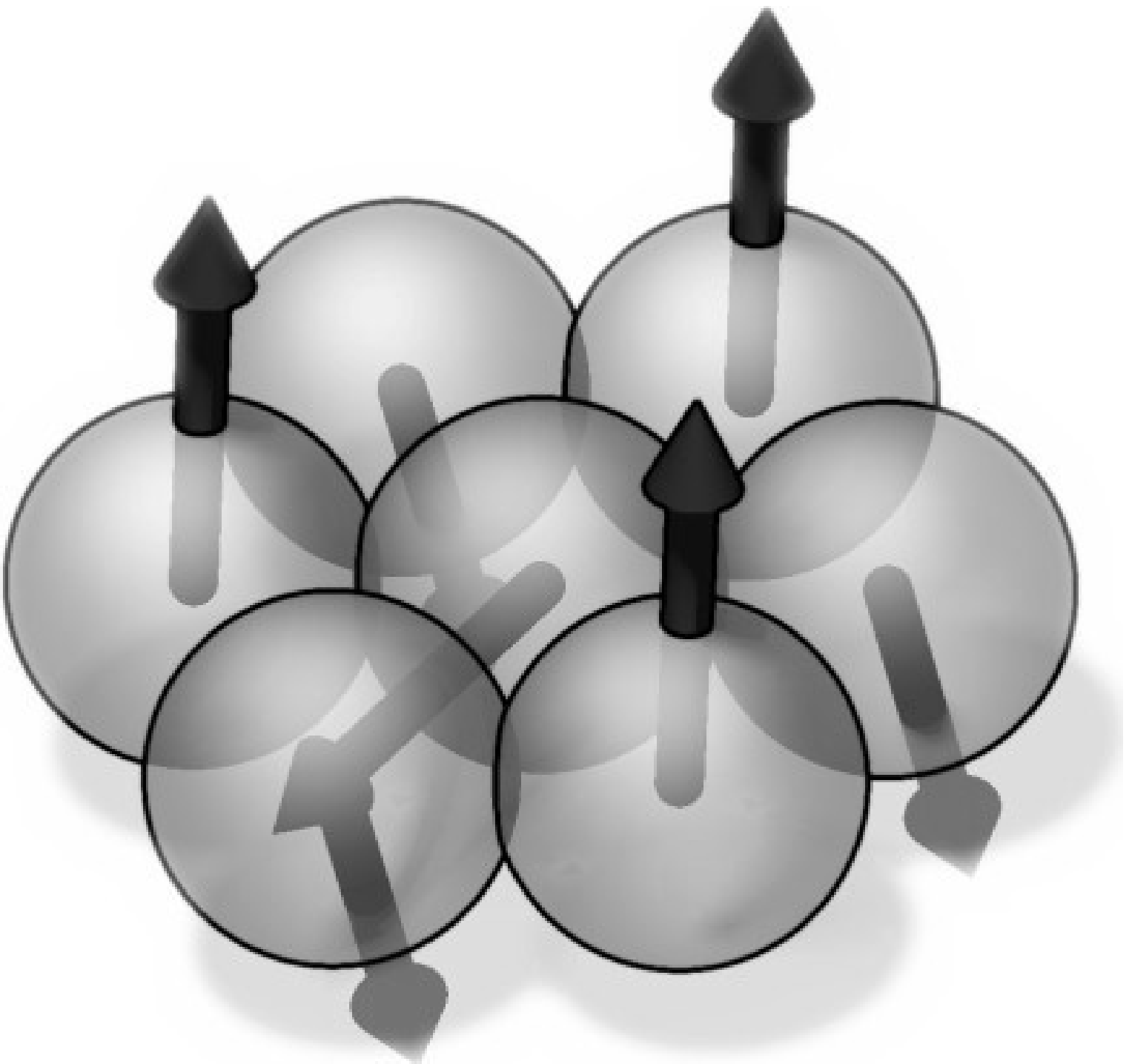}
\label{figcr1:d}}\\
\subfigure[]{
\includegraphics*[width=0.17\textwidth]{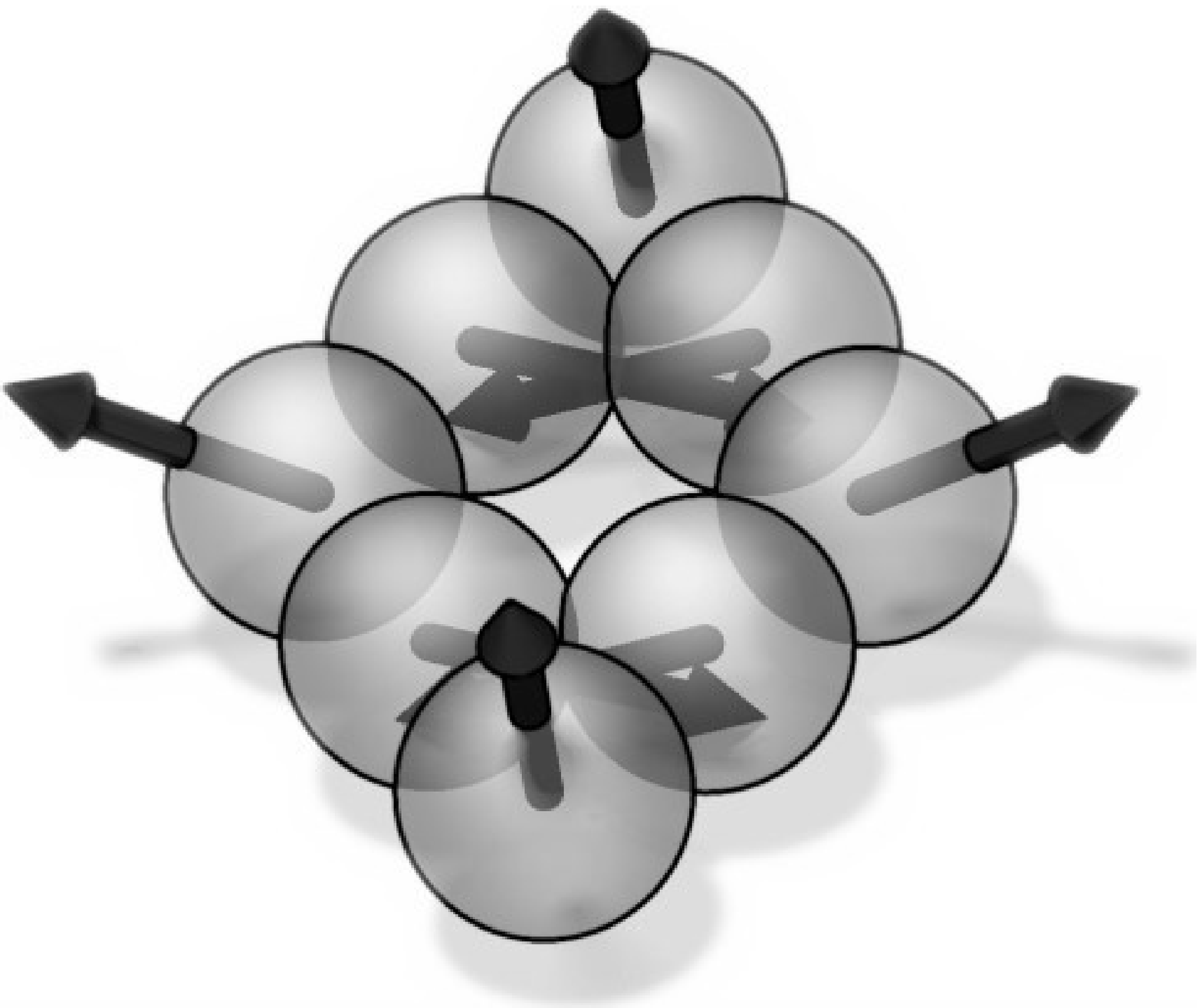}
\label{figcr1:e}}
\subfigure[]{
\includegraphics*[width=0.16\textwidth]{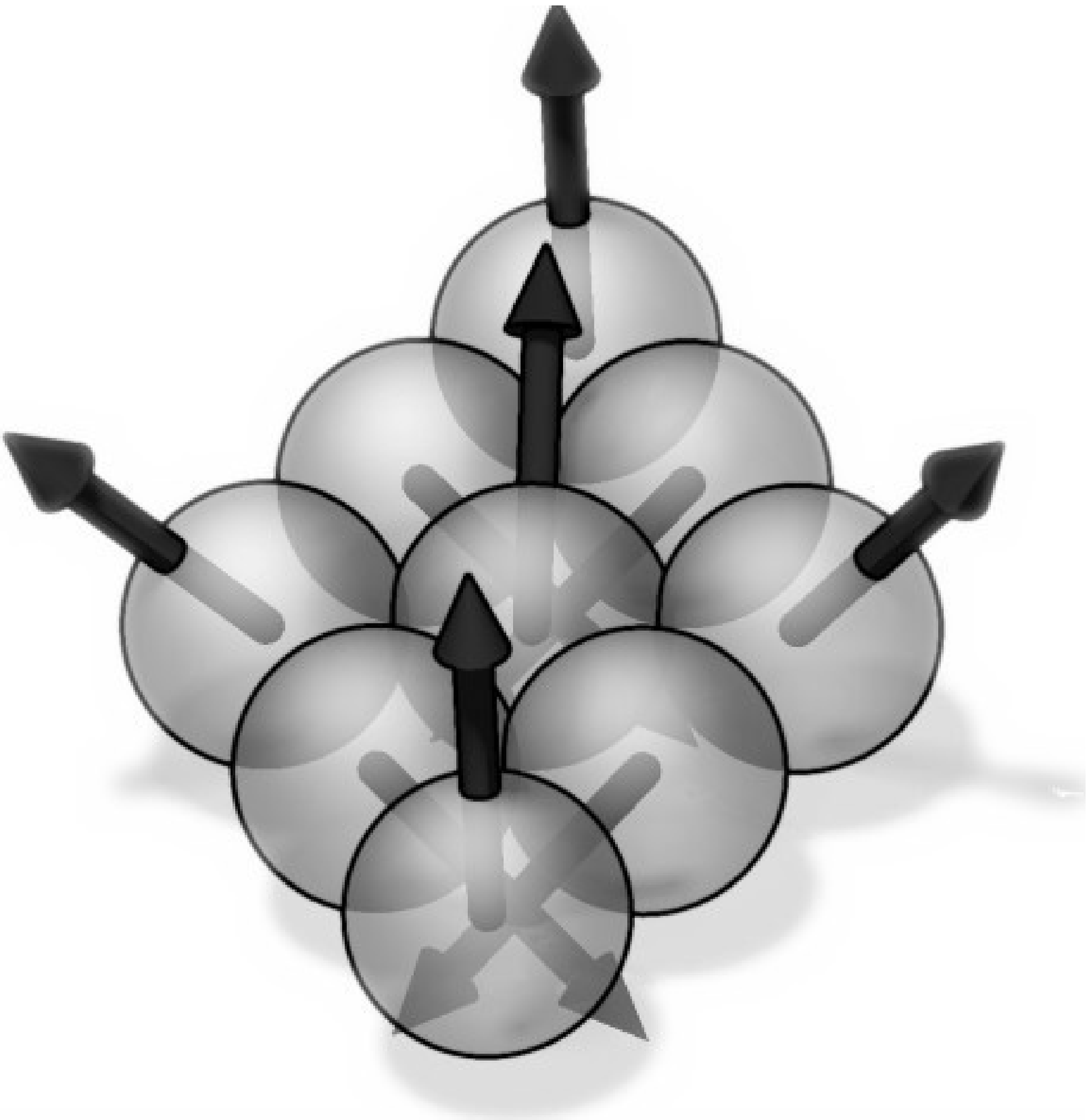}
\label{figcr1:f}}\\
\caption{\label{figcr1} The calculated magnetic ground state for Cr
clusters on a Cu(111) surface. } 
\end{center}
\end{figure}

In Fig.~\ref{figcr1} the geometries and calculated magnetic
configurations for a selection of Cr clusters are shown. Calculations
for Mn clusters with similar geometries can be found
elsewhere.\cite{Bergman06} The magnetic moments obtained for each atom
of the Cr clusters shown in Fig.~\ref{figcr1} can be seen in Table.~\ref{table:cr1}. 
The magnetic structures are for most of the clusters quite similar to
the calculated magnetic structure for Mn clusters although certain
differences occur, as will be commented on below. Both the Cr dimer
and a straight trimer (not shown) orders antiferromagnetically while a
three atom triangle has the same non-collinear structure as its Mn
counterpart, as shown in Fig.~\ref{figmnt1:a}. It can be noted that
this non-collinear structure has also been reported from calculations
on Cr clusters, with the same geometry, supported on
Au(111).\cite{gotsis2006,bergmanCr} 
\begin{table}
\caption{\label{table:cr1} Magnetic moments (in $\mu_B$) for atoms in
the clusters displayed in Fig.~\ref{figcr1:a}-\ref{figcr1:f}. The atoms are numbered
starting from the leftmost atom, counting around the cluster in the
clockwise direction and, if applicable, ending with the central atom. }
\begin{center}
\begin{ruledtabular}
\begin{tabular}{ccccccc}
 ~ & 5(a) & 5(b) & 5(c) & 5(d) & 5(e) & 5(f)  \\
\hline
1 & 3.99& 3.99&  3.82&  3.54&  3.87&  3.57\\
2 & 3.99& 3.89&  3.82&  3.54&  3.65&  3.28\\
3 & 3.99& 3.92&  3.82&  3.54&  3.89&  3.89\\
4 & 3.50& 3.89&  3.82&  3.54&  3.65&  3.28\\
5 &  -  & 3.99&  3.82&  3.54&  3.87&  3.57\\
6 &  -  &  -  &  3.82&  3.54&  3.65&  3.28\\
7 &  -  &  -  &   -  &  3.12&  3.89&  3.89\\
8 &  -  &  -  &   -  &   -  &  3.65&  3.28\\
9 &  -  &  -  &   -  &   -  &   -  &  2.81\\
\end{tabular}
\end{ruledtabular}
\end{center}
\end{table}
The cluster in Fig.~\ref{figcr1:a} has a collinear antiferromagnetic
ground state since the edge atoms only have one nearest neighbour and
therefore no frustration occurs. The pentamer in Fig.~\ref{figcr1:b}
also exhibits an antiferromagnetic ground-state, which in contrast to
the non-collinear behaviour of the Mn pentamer, is purely
collinear. This finding indicates that the magnetic structures of the
Cr clusters are more strongly dependant on the nearest-neighbour
exchange coupling than the Mn clusters are. The calculated exchange
parameters $J_{ij}$ for the Cr pentamer is shown in
Table~\ref{tab:table2}. Compared with the exchange interactions for
the Mn pentamer in Table~\ref{tab:table1} we see that the
nearest-neighbour interactions are indeed larger between the Cr
atoms. On the other hand, the more long ranged interactions also have
larger magnitudes in the Cr pentamer than for the Mn counterpart. The
exchange interactions between atoms further from each other do however
not always compete against the nearest-neighbour interactions as was
the case for the Mn pentamer. 
\begin{table}
\caption{\label{tab:table2}Calculated exchange parameters $J_{ij}$ (in
meV) for the Cr pentamer shown in Fig.~\ref{figcr1:b}. The atoms are
numbered from left to right.} 
\begin{ruledtabular}
\begin{tabular}{cccccc}
 $i \backslash j$ & 1 & 2 & 3 & 4 & 5\\
\hline
  1 & - & -143 & 4.6 & 18.9 & 15.3 \\
  2 & -143 & - & -97.0 & -40.5 & 18.9 \\
  3 & 4.6 & -97.0 & - & -97.0 & 4.6 \\
  4 & 18.9  & -40.5 & -97.0 & - & -143 \\
  5 & 15.3 & 18.9 & 4.6  & -143 & -\\
\end{tabular}
\end{ruledtabular}
\end{table}
\par
The collinear antiferromagnetic behaviour found for the pentamer also
occurs for the six atom ring displayed in Fig.~\ref{figcr1:c}, which
is expected since the geometry does not cause frustration for the
nearest-neighbour interactions. 
The cluster in Fig.~\ref{figcr1:d} has a symmetric non-collinear
ground state with an angle between two neighbouring atoms on the rim
of the cluster of 157$^\circ$ and between the central atom and any
outer atom the angle is 101$^\circ$. This can be compared with the
energy minimum obtained when minimizing the nearest-neighbour
Heisenberg Hamiltonian where all nearest neighbour $J_{ij}$'s are set
to be negative but equal. The Heisenberg model would give equilibrium
angles of the same cluster geometry of between 151$^\circ$ outer
neighbours and 104$^\circ$ between the central atom and any
neighbour. Although the agreement between our calculated ground state
and the Heisenberg minimum is good, it should be noted that due to the
difference in the local structure around the central and outer atoms
the exchange parameters $J_{ij}$ should be different between two outer
atoms compared to $J_{ij}$'s connecting to the central atom. This
difference can be considered in a simple model analysis by damping the
strength of the exchange parameters where the central atom is
connected, and for a damping of 20\% for these exchange parameters,
the Heisenberg Hamiltonian approach yields an energy minimum with the
angles of 157$^\circ$ for neighbouring outer atoms and 101$^\circ$
between the central atom and an edge atom which are in perfect
agreement with our calculated angles.  
\par
\begin{table}
\caption{\label{table:cr1e} Angles between magnetic moments for atoms in the
cluster displayed in Fig.~\ref{figcr1:e}. The atoms are numbered starting from the
leftmost atom and increasing along the clockwise direction of the cluster. }
\begin{center}
\begin{ruledtabular}
\begin{tabular}{cccccccccc}
 Atom & 1 & 2 & 3 & 4 & 5 & 6 & 7 & 8 \\
\hline
1 & 0 &169& 56& 56&112& 56& 56&169\\
2 &169& 0 &112&133& 56&133&112& 0 \\
3 & 56&112& 0 &112& 56&112& 0 &112\\
4 & 56&133&112& 0 &169& 0 &112&133\\
5 &112& 56& 56&169& 0 &169& 56& 56\\
6 & 56&133&112& 0 &169& 0 &112&133\\
7 & 56&112& 0 &112& 56&112& 0 &112\\
8 &169& 0 &112&133& 56&133&112& 0 \\
%\\
\end{tabular}
\end{ruledtabular}
\end{center}
\end{table}
Since the atoms of the hollow cluster in Fig.~\ref{figcr1:e} do not
all have only two nearest neighbours, as is the case for the other
ring like geometry of Fig.~\ref{figcr1:c}, the cluster can not have an
unfrustrated antiferromagnetic solution. Describing the cluster with a
Heisenberg Hamiltonian with equal and negative $J_{ij}$'s would yield
a ground state with 120$^\circ$ between neighbouring atoms. As seen in
Fig.~\ref{figcr1:e} our calculated magnetic structure differs from the
Heisenberg minimum, it is instead described with three different
nearest-neighbour angles. The upper and lower edge atoms have an angle
of 112$^\circ$ to their neighbours while the leftmost and rightmost of
the atoms have an angle of 169$^\circ$ to their nearest
neighbours. The third angle is that between two atoms with three
neighbours each and they have an angle of 133$^\circ$ between
them. This structure can be explained in a similar way as the cluster
in Fig.~\ref{figcr1:d} could, with different local geometries
resulting in different strengths of the exchange coupling and thus
different $J_{ij}$'s for different atoms. The angles between the
different magnetic moments for the cluster shown in
Fig.~\ref{figcr1:e} are given in Table.~\ref{table:cr1e}. 
\par
\begin{table}
\caption{\label{table:cr1f} Angles between magnetic moments for atoms
in the cluster displayed in Fig.~\ref{figcr1:f}. The atoms are
numbered starting from the leftmost atom, increasing along the
clockwise direction of the cluster and ending with the central atom.}
\begin{center}
\begin{ruledtabular}
\begin{tabular}{cccccccccc}
 Atom & 1 & 2 & 3 & 4 & 5 & 6 & 7 & 8 & 9 \\
\hline
1&    0&  156&   40&   95&   74&   95&   40&  156&   69\\
2&  156&    0&  137&   86&   95&   86&  137&    0&  128\\
3&   40&  137&    0&  137&   40&  137&    0&  137&   40\\
4&   95&   86&  137&    0&  156&    0&  137&   86&  128\\
5&   74&   95&   40&  156&    0&  156&   40&   95&   69\\
6&   95&   86&  137&    0&  156&    0&  137&   86&  128\\
7&   40&  137&    0&  137&   40&  137&    0&  137&   40\\
8&  156&    0&  137&   86&   95&   86&  137&    0&  128\\
9&   69&  128&   40&  128&   69&  128&   40&  128&    0\\
%\\
\end{tabular}
\end{ruledtabular}
\end{center}
\end{table}
Filling the empty site in the middle of the cluster in
Fig.~\ref{figcr1:e} gives the cluster geometry shown in
Fig.~\ref{figcr1:f}. This additional atom causes the magnetic
structure to be even more complex. Since the symmetry is lowered
compared to the cluster in Fig.~\ref{figcr1:d}, the central atom does
not have the same angle towards all of its neighbours. Instead two
angles are needed to describe the structure of the neighbours of the
central atom. One of these is the angle of 69$^\circ$ which the
central atom makes towards the leftmost and rightmost atoms and the
other angle connects the central atom with the remaining four
neighbours and the size of this angle is 128$^\circ$. The upper and
lower edge atoms makes and angle of 137$^\circ$ with their
neighbours.  The angles between different magnetic moments for the cluster shown in
Fig.~\ref{figcr1:f} are given in Table.~\ref{table:cr1f}. 
\subsection{Three dimensional clusters}
\begin{figure}
\begin{center}
\subfigure[]{
\includegraphics*[width=0.13\textwidth]{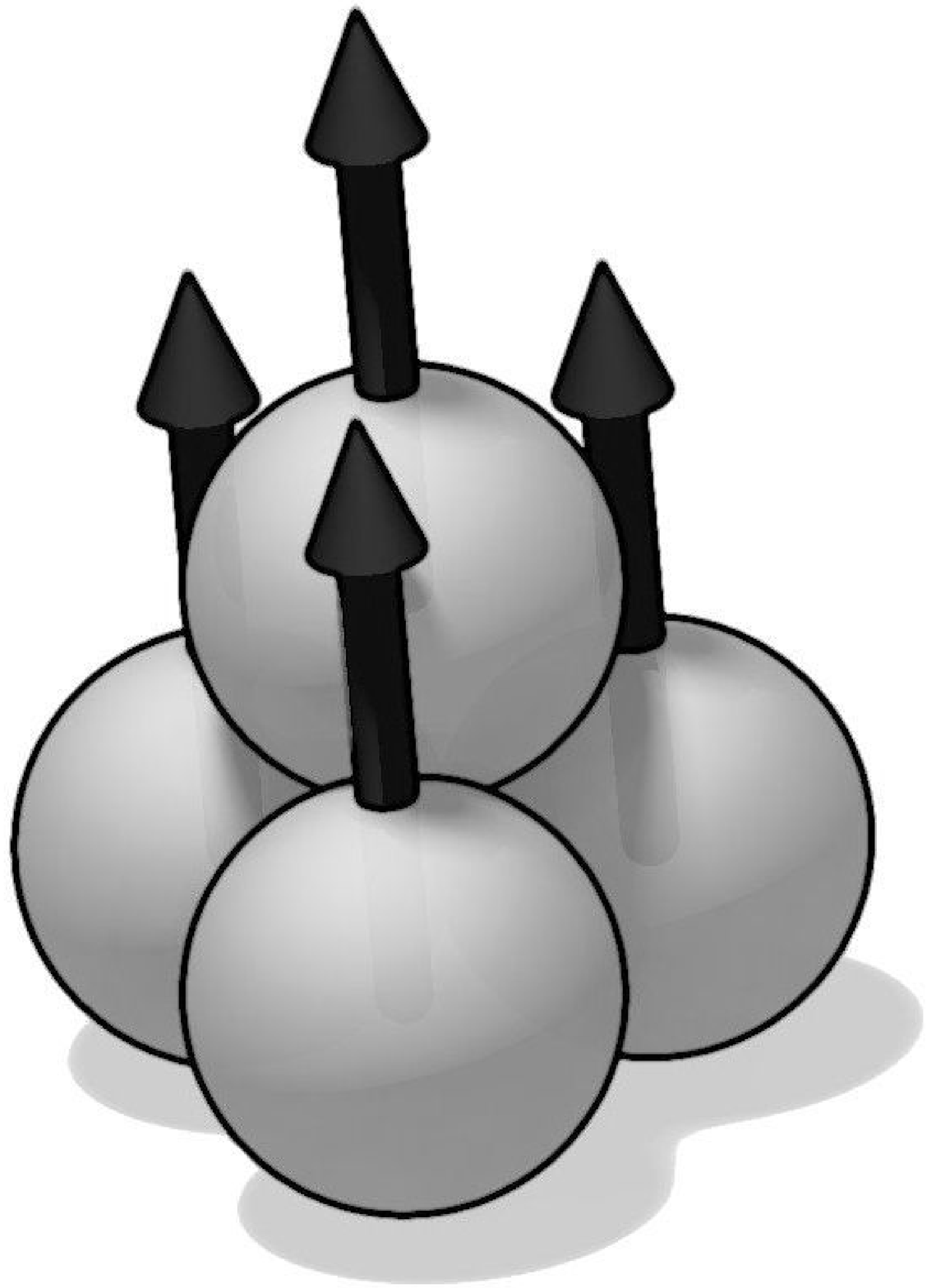}
\label{figpyra:a}}
\subfigure[]{
\includegraphics*[width=0.15\textwidth]{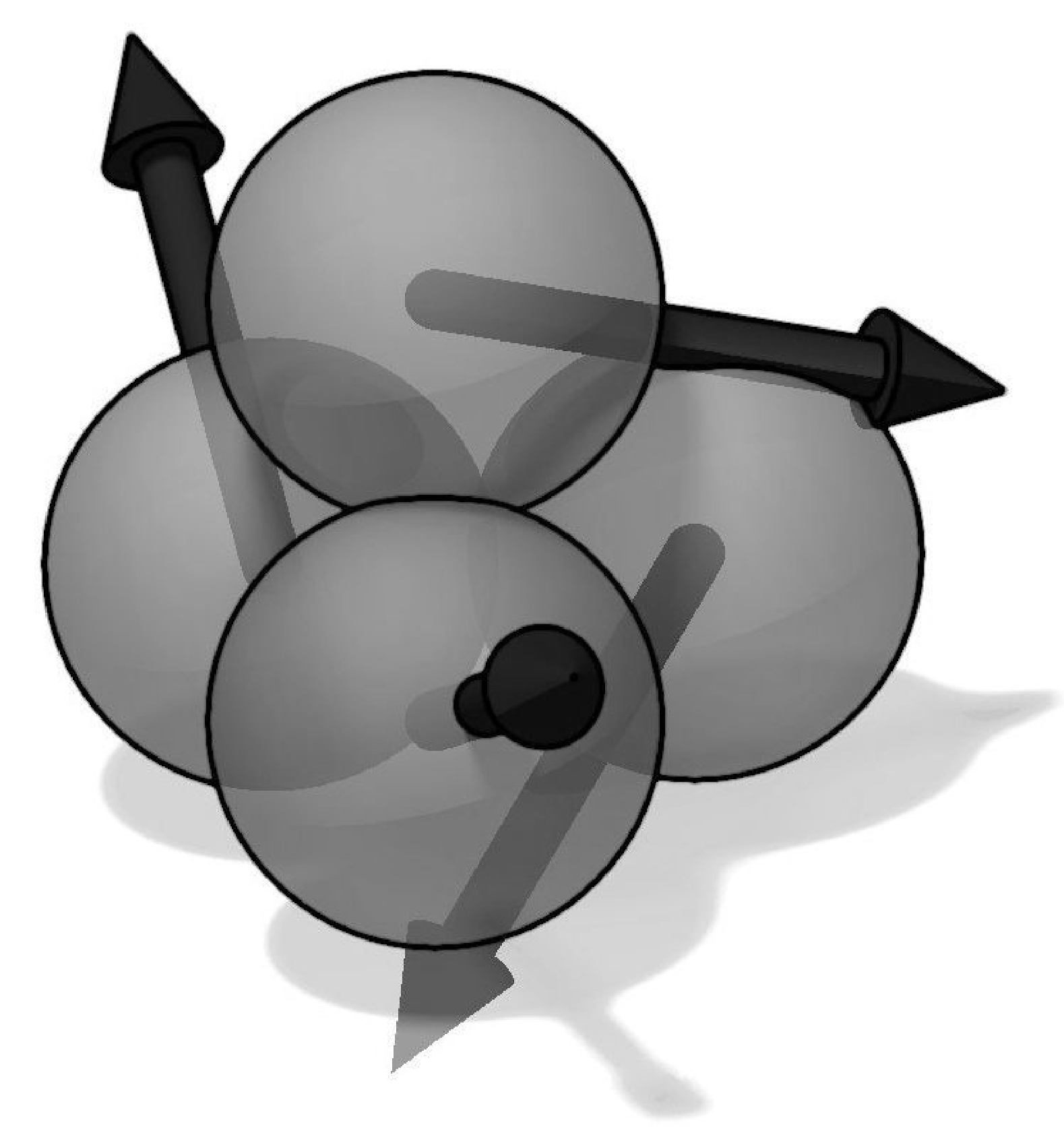}
\label{figpyra:b}}
\subfigure[]{
\includegraphics*[width=0.14\textwidth]{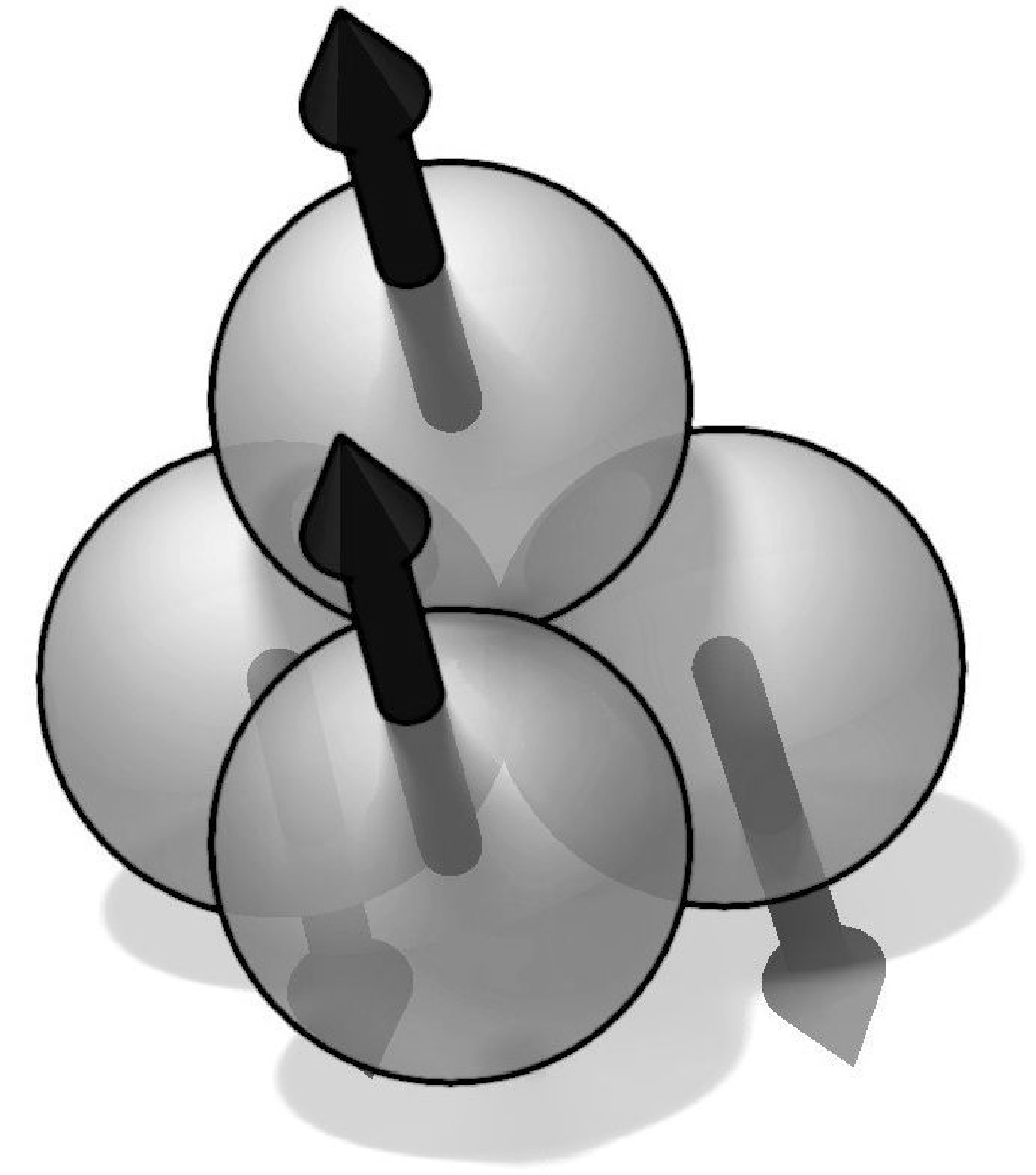}
\label{figpyra:c}}
\caption{\label{figpyra} The calculated magnetic ordering for pyramid
shaped clusters on a Cu(111) surface. Fig.~\ref{figpyra:a} shows a Fe
cluster with a ferromagnetic solution. Fig.~\ref{figpyra:b} shows a Mn
cluster and Fig.~\ref{figpyra:c} shows the Cr pyramid.} 
\end{center} 
\end{figure}

So far the studied clusters have all been confined in one layer above
the Cu surface. However, our method can treat three dimensional
clusters as well. In order to demonstrate this we show the obtained
magnetic configurations for a pyramid-like tetrahedron shaped cluster
in Fig.~\ref{figpyra}. 
As expected from the results for Fe clusters reported earlier in this
work, the Fe cluster (Fig.~\ref{figpyra:a}) exhibits a ferromagnetic
order. The atom situated on top of the pyramid has a magnetic moment
of 3.40\uB while the three Fe atoms closer to the Cu surface have a
magnetic moment of 3.11\uB. For the Mn cluster shown in
Fig.~\ref{figpyra:b}, a non-collinear structure is found and for the
Cr pyramid, shown in Fig.~\ref{figpyra:c} a collinear
antiferromagnetic solution is found. A model Heisenberg Hamiltonian,
as in Eqn.~\ref{eqn4}, with only antiferromagnetic nearest-neighbour
exchange parameters, $J_{ij}$, of equal size, yields a two-fold
degenerate ground state, either a collinear antiferromagnet or a
non-collinear tetragonal configuration with 109$^{\circ}$ between
neighbouring angles.  The calculation for the Mn pyramid over Cu(111)
shows angles which are slightly distorted relative to those of the
free pyramid, around 116 degrees between the base atoms close to the
substrate and angles of about 100 degrees between the base site and
the top one. It is peculiar that the Mn cluster has the non-collinear
tetragonal configuration as the ground state whereas the Cr cluster
has the collinear antiferromagnetic ground state solution. A possible
explanation for the non-collinear ground state of the Mn pyramid could
be that, similar to the situation for several of the planar clusters,
the reduced neighbour coordination for the top atom compared to the
atom in the base of the pyramid yields different exchange interaction
strengths between the atoms in the cluster. The calculated exchange
parameters confirm this since the exchange interaction between a top
and a base atom is $-83$ meV while the interaction between two
base atoms $-46$ meV. The reduced neighbour coordination also affects
the magnetic moment for the top atom which is 4.5~$\mu_B$ compared to
4.0~$\mu_B$ for the base atoms.
However, the situation with different exchange parameters is also
present for the Cr pyramid where the interaction strength
between the top atom and a base atom is $-171$ meV compared to $-98$
meV for the coupling between two base atoms. The magnetic moment for
the Cr atom on top of the pyramid is 4.2~$\mu_B$ while the magnetic
moment for the base atoms is 3.6~$\mu_B$.
The difference in the magnetic ordering found for the Cr and Mn
pyramids indicates that the bilinear exchange terms $J$ cannot always
describe the magnetic interactions between the atoms in supported
magnetic clusters, a fact which previously been suggested for magnetic
dimers on surfaces.\cite{Costa04} It can be noted that the difference
in the total energy between the non-collinear ground state structure
and the antiferromagnetic solution for the Mn cluster is 25 meV per
atom while the corresponding difference for the Cr pyramid is -15 meV
per atom. 
\section{Conclusions}
We have studied the magnetic structure of small clusters of Fe, Mn,
and Cr supported on a Cu(111) surface with non-collinear, first
principles calculations. The studied Fe clusters are found to order
ferromagnetically regardless of the cluster geometry. For Mn and Cr
clusters, antiferromagnetic exchange interactions between
nearest-neighbours are found to cause either collinear
antiferromagnetic ordering or non-collinear ordering. The
non-collinear ordering occurs when the cluster geometry is such that
an antiferromagnetic arrangement becomes frustrated. The calculations
have been accompanied by comparisons with calculated effective exchange
interactions as well as with ground states obtained from a
simplified Heisenberg Hamiltonian and the comparisons show that the
exchange interactions vary for different atoms in the clusters as a
result of the different local structure. Differences between the
magnetic ordering for Mn and Cr clusters are found where Cr clusters
seem to prefer collinear solutions to a higher degree while Mn
clusters can exhibit non-collinear configurations even for
unfrustrated cluster geometries. Comparisons with model Hamiltonians
show that the magnetic structure of certain clusters can be explained
by a simple nearest-neighbour Heisenberg Hamiltonian while other
cluster geometries cause more complex behaviours. 
\section{Acknowledgements}
We acknowledge financial support from the G\"oran Gustafsson
foundation, the Swedish Research Council, the Swedish Foundation for
Strategic Research, Lennanders Stiftelse, and  CNPq, Brazil. The
calculations were performed at the high performance computing centers
UPPMAX, and PDC within the Swedish National Infrastructure for
Computing and at the computational facilities of the LCCA, University
of S\~ao Paulo and of the CENAPAD at University of Campinas, SP,
Brazil.


\begin{thebibliography}{60}
\expandafter\ifx\csname natexlab\endcsname\relax\def\natexlab#1{#1}\fi
\expandafter\ifx\csname bibnamefont\endcsname\relax
  \def\bibnamefont#1{#1}\fi
\expandafter\ifx\csname bibfnamefont\endcsname\relax
  \def\bibfnamefont#1{#1}\fi
\expandafter\ifx\csname citenamefont\endcsname\relax
  \def\citenamefont#1{#1}\fi
\expandafter\ifx\csname url\endcsname\relax
  \def\url#1{\texttt{#1}}\fi
\expandafter\ifx\csname urlprefix\endcsname\relax\def\urlprefix{URL }\fi
\providecommand{\bibinfo}[2]{#2}
\providecommand{\eprint}[2][]{\url{#2}}

\bibitem[{\citenamefont{Binnig and Rohrer}(1982)}]{Binnig}
\bibinfo{author}{\bibfnamefont{G.}~\bibnamefont{Binnig}} \bibnamefont{and}
  \bibinfo{author}{\bibfnamefont{H.}~\bibnamefont{Rohrer}},
  \bibinfo{journal}{Helv. Phys. Acta.} \textbf{\bibinfo{volume}{55}},
  \bibinfo{pages}{726} (\bibinfo{year}{1982}).

\bibitem[{\citenamefont{Himpsel et~al.}(1998)\citenamefont{Himpsel, Ortega,
  Mankey, and Willis}}]{Himpsel}
\bibinfo{author}{\bibfnamefont{F.~J.} \bibnamefont{Himpsel}},
  \bibinfo{author}{\bibfnamefont{J.~E.} \bibnamefont{Ortega}},
  \bibinfo{author}{\bibfnamefont{G.~J.} \bibnamefont{Mankey}},
  \bibnamefont{and} \bibinfo{author}{\bibfnamefont{R.~F.}
  \bibnamefont{Willis}}, \bibinfo{journal}{Adv. Phys.}
  \textbf{\bibinfo{volume}{47}}, \bibinfo{pages}{511} (\bibinfo{year}{1998}).

\bibitem[{\citenamefont{Gambardella et~al.}(2001)\citenamefont{Gambardella,
  Blanc, Kuhnke, Kern, Picaud, Ramseyer, Girardet, Barreteau, Spanjaard, and
  Desjonqu\`eres}}]{Gambardella2001}
\bibinfo{author}{\bibfnamefont{P.}~\bibnamefont{Gambardella}},
  \bibinfo{author}{\bibfnamefont{M.}~\bibnamefont{Blanc}},
  \bibinfo{author}{\bibfnamefont{K.}~\bibnamefont{Kuhnke}},
  \bibinfo{author}{\bibfnamefont{K.}~\bibnamefont{Kern}},
  \bibinfo{author}{\bibfnamefont{F.}~\bibnamefont{Picaud}},
  \bibinfo{author}{\bibfnamefont{C.}~\bibnamefont{Ramseyer}},
  \bibinfo{author}{\bibfnamefont{C.}~\bibnamefont{Girardet}},
  \bibinfo{author}{\bibfnamefont{C.}~\bibnamefont{Barreteau}},
  \bibinfo{author}{\bibfnamefont{D.}~\bibnamefont{Spanjaard}},
  \bibnamefont{and} \bibinfo{author}{\bibfnamefont{M.~C.}
  \bibnamefont{Desjonqu\`eres}}, \bibinfo{journal}{Phys. Rev. B}
  \textbf{\bibinfo{volume}{64}}, \bibinfo{pages}{045404}
  (\bibinfo{year}{2001}).

\bibitem[{\citenamefont{Tsunoda}(1989)}]{Tsunoda89}
\bibinfo{author}{\bibfnamefont{Y.}~\bibnamefont{Tsunoda}},
  \bibinfo{journal}{Journal of Physics: Condensed Matter}
  \textbf{\bibinfo{volume}{1}}, \bibinfo{pages}{10427} (\bibinfo{year}{1989}).

\bibitem[{\citenamefont{Sj\"ostedt and Nordstr\"om}(2002)}]{Sjostedt02}
\bibinfo{author}{\bibfnamefont{E.}~\bibnamefont{Sj\"ostedt}} \bibnamefont{and}
  \bibinfo{author}{\bibfnamefont{L.}~\bibnamefont{Nordstr\"om}},
  \bibinfo{journal}{Phys. Rev. B} \textbf{\bibinfo{volume}{66}},
  \bibinfo{eid}{014447} (\bibinfo{year}{2002}).

\bibitem[{\citenamefont{Fawcett}(1988)}]{Fawcett88}
\bibinfo{author}{\bibfnamefont{E.}~\bibnamefont{Fawcett}},
  \bibinfo{journal}{Rev. Mod. Phys.} \textbf{\bibinfo{volume}{60}},
  \bibinfo{pages}{209} (\bibinfo{year}{1988}).

\bibitem[{\citenamefont{Bodeker et~al.}(1999)\citenamefont{Bodeker, Schreyer,
  and Zabel}}]{Bodeker99}
\bibinfo{author}{\bibfnamefont{P.}~\bibnamefont{Bodeker}},
  \bibinfo{author}{\bibfnamefont{A.}~\bibnamefont{Schreyer}}, \bibnamefont{and}
  \bibinfo{author}{\bibfnamefont{H.}~\bibnamefont{Zabel}},
  \bibinfo{journal}{Phys. Rev. B} \textbf{\bibinfo{volume}{59}},
  \bibinfo{pages}{9408} (\bibinfo{year}{1999}).

\bibitem[{\citenamefont{Robles et~al.}(2003)\citenamefont{Robles, Martinez,
  Stoeffler, and Vega}}]{Robles03}
\bibinfo{author}{\bibfnamefont{R.}~\bibnamefont{Robles}},
  \bibinfo{author}{\bibfnamefont{E.}~\bibnamefont{Martinez}},
  \bibinfo{author}{\bibfnamefont{D.}~\bibnamefont{Stoeffler}},
  \bibnamefont{and} \bibinfo{author}{\bibfnamefont{A.}~\bibnamefont{Vega}},
  \bibinfo{journal}{Phys. Rev. B} \textbf{\bibinfo{volume}{68}},
  \bibinfo{eid}{094413} (\bibinfo{year}{2003}).

\bibitem[{\citenamefont{Bradley and Thewlis}(1927)}]{Bradley27}
\bibinfo{author}{\bibfnamefont{A.}~\bibnamefont{Bradley}} \bibnamefont{and}
  \bibinfo{author}{\bibfnamefont{J.}~\bibnamefont{Thewlis}},
  \bibinfo{journal}{Proc. R. Soc. London} \textbf{\bibinfo{volume}{115}},
  \bibinfo{pages}{465} (\bibinfo{year}{1927}).

\bibitem[{\citenamefont{Hobbs et~al.}(2003)\citenamefont{Hobbs, Hafner, and
  Spisak}}]{Hobbs03}
\bibinfo{author}{\bibfnamefont{D.}~\bibnamefont{Hobbs}},
  \bibinfo{author}{\bibfnamefont{J.}~\bibnamefont{Hafner}}, \bibnamefont{and}
  \bibinfo{author}{\bibfnamefont{D.}~\bibnamefont{Spisak}},
  \bibinfo{journal}{Phys. Rev. B} \textbf{\bibinfo{volume}{68}},
  \bibinfo{eid}{014407} (\bibinfo{year}{2003}).

\bibitem[{\citenamefont{Eriksson et~al.}(2004)\citenamefont{Eriksson,
  Lizarraga, Felton, Bergqvist, Andersson, Nordblad, and
  Eriksson}}]{teriksson04}
\bibinfo{author}{\bibfnamefont{T.}~\bibnamefont{Eriksson}},
  \bibinfo{author}{\bibfnamefont{R.}~\bibnamefont{Lizarraga}},
  \bibinfo{author}{\bibfnamefont{S.}~\bibnamefont{Felton}},
  \bibinfo{author}{\bibfnamefont{L.}~\bibnamefont{Bergqvist}},
  \bibinfo{author}{\bibfnamefont{Y.}~\bibnamefont{Andersson}},
  \bibinfo{author}{\bibfnamefont{P.}~\bibnamefont{Nordblad}}, \bibnamefont{and}
  \bibinfo{author}{\bibfnamefont{O.}~\bibnamefont{Eriksson}},
  \bibinfo{journal}{Phys. Rev. B} \textbf{\bibinfo{volume}{69}},
  \bibinfo{pages}{054422} (\bibinfo{year}{2004}).

\bibitem[{\citenamefont{Eriksson et~al.}(2005)\citenamefont{Eriksson,
  Bergqvist, Andersson, Nordblad, and Eriksson}}]{teriksson05}
\bibinfo{author}{\bibfnamefont{T.}~\bibnamefont{Eriksson}},
  \bibinfo{author}{\bibfnamefont{L.}~\bibnamefont{Bergqvist}},
  \bibinfo{author}{\bibfnamefont{Y.}~\bibnamefont{Andersson}},
  \bibinfo{author}{\bibfnamefont{P.}~\bibnamefont{Nordblad}}, \bibnamefont{and}
  \bibinfo{author}{\bibfnamefont{O.}~\bibnamefont{Eriksson}},
  \bibinfo{journal}{Phys. Rev. B} \textbf{\bibinfo{volume}{72}},
  \bibinfo{pages}{144427} (\bibinfo{year}{2005}).

\bibitem[{\citenamefont{Billas et~al.}(1993)\citenamefont{Billas, Becker,
  Chatelain, and de~Heer}}]{Billas93}
\bibinfo{author}{\bibfnamefont{I.~M.~L.} \bibnamefont{Billas}},
  \bibinfo{author}{\bibfnamefont{J.~A.} \bibnamefont{Becker}},
  \bibinfo{author}{\bibfnamefont{A.}~\bibnamefont{Chatelain}},
  \bibnamefont{and} \bibinfo{author}{\bibfnamefont{W.~A.}
  \bibnamefont{de~Heer}}, \bibinfo{journal}{Phys. Rev. Lett.}
  \textbf{\bibinfo{volume}{71}}, \bibinfo{pages}{4067} (\bibinfo{year}{1993}).

\bibitem[{\citenamefont{Knickelbein}(2001)}]{Knickelbein01}
\bibinfo{author}{\bibfnamefont{M.~B.} \bibnamefont{Knickelbein}},
  \bibinfo{journal}{Phys. Rev. Lett.} \textbf{\bibinfo{volume}{86}},
  \bibinfo{pages}{5255} (\bibinfo{year}{2001}).

\bibitem[{\citenamefont{Douglass et~al.}(1992)\citenamefont{Douglass, Bucher,
  and Bloomfield}}]{Douglass92}
\bibinfo{author}{\bibfnamefont{D.~C.} \bibnamefont{Douglass}},
  \bibinfo{author}{\bibfnamefont{J.~P.} \bibnamefont{Bucher}},
  \bibnamefont{and} \bibinfo{author}{\bibfnamefont{L.~A.}
  \bibnamefont{Bloomfield}}, \bibinfo{journal}{Phys. Rev. B}
  \textbf{\bibinfo{volume}{45}}, \bibinfo{pages}{6341} (\bibinfo{year}{1992}).

\bibitem[{\citenamefont{Kohl and Bertsch}(1999)}]{Kohl99}
\bibinfo{author}{\bibfnamefont{C.}~\bibnamefont{Kohl}} \bibnamefont{and}
  \bibinfo{author}{\bibfnamefont{G.~F.} \bibnamefont{Bertsch}},
  \bibinfo{journal}{Phys. Rev. B} \textbf{\bibinfo{volume}{60}},
  \bibinfo{pages}{4205} (\bibinfo{year}{1999}).

\bibitem[{\citenamefont{Morisato et~al.}(2005)\citenamefont{Morisato, Khanna,
  and Kawazoe}}]{Morisato05}
\bibinfo{author}{\bibfnamefont{T.}~\bibnamefont{Morisato}},
  \bibinfo{author}{\bibfnamefont{S.~N.} \bibnamefont{Khanna}},
  \bibnamefont{and} \bibinfo{author}{\bibfnamefont{Y.}~\bibnamefont{Kawazoe}},
  \bibinfo{journal}{Phys. Rev. B} \textbf{\bibinfo{volume}{72}},
  \bibinfo{eid}{014435} (\bibinfo{year}{2005}).

\bibitem[{\citenamefont{Longo et~al.}(2005)\citenamefont{Longo, Noya, and
  Gallego}}]{Longo05}
\bibinfo{author}{\bibfnamefont{R.~C.} \bibnamefont{Longo}},
  \bibinfo{author}{\bibfnamefont{E.~G.} \bibnamefont{Noya}}, \bibnamefont{and}
  \bibinfo{author}{\bibfnamefont{L.~J.} \bibnamefont{Gallego}},
  \bibinfo{journal}{Phys. Rev. B} \textbf{\bibinfo{volume}{72}},
  \bibinfo{eid}{174409} (\bibinfo{year}{2005}).

\bibitem[{\citenamefont{Lau et~al.}(2002)\citenamefont{Lau, F\"olisch,
  Nietubyc, Reif, and Wurth}}]{Lau02}
\bibinfo{author}{\bibfnamefont{J.~T.} \bibnamefont{Lau}},
  \bibinfo{author}{\bibfnamefont{A.}~\bibnamefont{F\"ohlisch}},
  \bibinfo{author}{\bibfnamefont{R.}~\bibnamefont{Nietubyc}},
  \bibinfo{author}{\bibfnamefont{M.}~\bibnamefont{Reif}}, \bibnamefont{and}
  \bibinfo{author}{\bibfnamefont{W.}~\bibnamefont{Wurth}},
  \bibinfo{journal}{Phys. Rev. Lett.} \textbf{\bibinfo{volume}{89}},
  \bibinfo{eid}{057201} (\bibinfo{year}{2002}).

\bibitem[{\citenamefont{Binns et~al.}(2001)\citenamefont{Binns, Edmonds, Baker,
  Thornton, and Upward}}]{Binns2001}
\bibinfo{author}{\bibfnamefont{C.}~\bibnamefont{Binns}},
  \bibinfo{author}{\bibfnamefont{K.~W.} \bibnamefont{Edmonds}},
  \bibinfo{author}{\bibfnamefont{S.~H.} \bibnamefont{Baker}},
  \bibinfo{author}{\bibfnamefont{S.~C.} \bibnamefont{Thornton}},
  \bibnamefont{and} \bibinfo{author}{\bibfnamefont{M.~D.}
  \bibnamefont{Upward}}, \bibinfo{journal}{Scripta Mater.}
  \textbf{\bibinfo{volume}{44}}, \bibinfo{pages}{1303} (\bibinfo{year}{2001}).

\bibitem[{\citenamefont{Spisak and Hafner}(2002)}]{Spisak02}
\bibinfo{author}{\bibfnamefont{D.}~\bibnamefont{Spisak}} \bibnamefont{and}
  \bibinfo{author}{\bibfnamefont{J.}~\bibnamefont{Hafner}},
  \bibinfo{journal}{Phys. Rev. B} \textbf{\bibinfo{volume}{65}},
  \bibinfo{eid}{235405} (\bibinfo{year}{2002}).

\bibitem[{\citenamefont{Lazarovits et~al.}(2003)\citenamefont{Lazarovits,
  Szunyogh, Weinberger, and Ujfalussy}}]{Lazarovits03}
\bibinfo{author}{\bibfnamefont{B.}~\bibnamefont{Lazarovits}},
  \bibinfo{author}{\bibfnamefont{L.}~\bibnamefont{Szunyogh}},
  \bibinfo{author}{\bibfnamefont{P.}~\bibnamefont{Weinberger}},
  \bibnamefont{and}
  \bibinfo{author}{\bibfnamefont{B.}~\bibnamefont{Ujfalussy}},
  \bibinfo{journal}{Phys. Rev. B} \textbf{\bibinfo{volume}{68}},
  \bibinfo{eid}{024433} (\bibinfo{year}{2003}).

\bibitem[{\citenamefont{Stepanyuk
  et~al.}(1997{\natexlab{a}})\citenamefont{Stepanyuk, Hergert, Rennert,
  Wildberger, Zeller, and Dederichs}}]{Stepanyuk97}
\bibinfo{author}{\bibfnamefont{V.~S.} \bibnamefont{Stepanyuk}},
  \bibinfo{author}{\bibfnamefont{W.}~\bibnamefont{Hergert}},
  \bibinfo{author}{\bibfnamefont{P.}~\bibnamefont{Rennert}},
  \bibinfo{author}{\bibfnamefont{K.}~\bibnamefont{Wildberger}},
  \bibinfo{author}{\bibfnamefont{R.}~\bibnamefont{Zeller}}, \bibnamefont{and}
  \bibinfo{author}{\bibfnamefont{P.~H.} \bibnamefont{Dederichs}},
  \bibinfo{journal}{J. Magn. Magn. Matter} \textbf{\bibinfo{volume}{165}},
  \bibinfo{pages}{272} (\bibinfo{year}{1997}{\natexlab{a}}).

\bibitem[{\citenamefont{Stepanyuk
  et~al.}(1997{\natexlab{b}})\citenamefont{Stepanyuk, Hergert, Wildberger,
  Nayak, and Jena}}]{Stepanyuk97_2}
\bibinfo{author}{\bibfnamefont{V.~S.} \bibnamefont{Stepanyuk}},
  \bibinfo{author}{\bibfnamefont{W.}~\bibnamefont{Hergert}},
  \bibinfo{author}{\bibfnamefont{K.}~\bibnamefont{Wildberger}},
  \bibinfo{author}{\bibfnamefont{S.~K.} \bibnamefont{Nayak}}, \bibnamefont{and}
  \bibinfo{author}{\bibfnamefont{P.}~\bibnamefont{Jena}},
  \bibinfo{journal}{Surf. Sci.} \textbf{\bibinfo{volume}{384}},
  \bibinfo{pages}{L892} (\bibinfo{year}{1997}{\natexlab{b}}).

\bibitem[{\citenamefont{Stepanyuk et~al.}(1998)\citenamefont{Stepanyuk,
  Hergert, Rennert, Kokko, Tatarchenko, and Wildberger}}]{Stepanyuk98}
\bibinfo{author}{\bibfnamefont{V.~S.} \bibnamefont{Stepanyuk}},
  \bibinfo{author}{\bibfnamefont{W.}~\bibnamefont{Hergert}},
  \bibinfo{author}{\bibfnamefont{P.}~\bibnamefont{Rennert}},
  \bibinfo{author}{\bibfnamefont{K.}~\bibnamefont{Kokko}},
  \bibinfo{author}{\bibfnamefont{A.~F.} \bibnamefont{Tatarchenko}},
  \bibnamefont{and}
  \bibinfo{author}{\bibfnamefont{K.}~\bibnamefont{Wildberger}},
  \bibinfo{journal}{Phys. Rev. B} \textbf{\bibinfo{volume}{57}},
  \bibinfo{pages}{15585} (\bibinfo{year}{1998}).

\bibitem[{\citenamefont{S.~Uzdin and Demangeat}(1999)}]{Uzdin1999}
\bibinfo{author}{\bibfnamefont{V.~U.} \bibnamefont{S.~Uzdin}} \bibnamefont{and}
  \bibinfo{author}{\bibfnamefont{C.}~\bibnamefont{Demangeat}},
  \bibinfo{journal}{Europhys. Lett.} \textbf{\bibinfo{volume}{47}},
  \bibinfo{pages}{556} (\bibinfo{year}{1999}).

\bibitem[{\citenamefont{S.~Uzdin and Demangeat}(2001)}]{Uzdin2001}
\bibinfo{author}{\bibfnamefont{V.~U.} \bibnamefont{S.~Uzdin}} \bibnamefont{and}
  \bibinfo{author}{\bibfnamefont{C.}~\bibnamefont{Demangeat}},
  \bibinfo{journal}{Surf. Sci.} \textbf{\bibinfo{volume}{482}},
  \bibinfo{pages}{965} (\bibinfo{year}{2001}).

\bibitem[{\citenamefont{Lounis et~al.}(2005)\citenamefont{Lounis, Mavropoulos,
  Dederichs, and Bl\"ugel}}]{Lounis05}
\bibinfo{author}{\bibfnamefont{S.}~\bibnamefont{Lounis}},
  \bibinfo{author}{\bibfnamefont{P.}~\bibnamefont{Mavropoulos}},
  \bibinfo{author}{\bibfnamefont{P.~H.} \bibnamefont{Dederichs}},
  \bibnamefont{and} \bibinfo{author}{\bibfnamefont{S.}~\bibnamefont{Bl\"ugel}},
  \bibinfo{journal}{Phys. Rev. B} \textbf{\bibinfo{volume}{72}},
  \bibinfo{eid}{224437} (\bibinfo{year}{2005}).

\bibitem[{\citenamefont{Robles and Nordstr\"{o}m}(2006)}]{Robles}
\bibinfo{author}{\bibfnamefont{R.}~\bibnamefont{Robles}} \bibnamefont{and}
  \bibinfo{author}{\bibfnamefont{L.}~\bibnamefont{Nordstr\"{o}m}},
  \bibinfo{journal}{Phys. Rev. B} \textbf{\bibinfo{volume}{74}},
  \bibinfo{pages}{094403} (\bibinfo{year}{2006}).

\bibitem[{\citenamefont{Lounis et~al.}()\citenamefont{Lounis, Reif,
  Mavropoulos, Glaser, Dederichs, Martins, Bl\"ugel, and Wurth}}]{Lounis06}
\bibinfo{author}{\bibfnamefont{S.}~\bibnamefont{Lounis}},
  \bibinfo{author}{\bibfnamefont{M.}~\bibnamefont{Reif}},
  \bibinfo{author}{\bibfnamefont{P.}~\bibnamefont{Mavropoulos}},
  \bibinfo{author}{\bibfnamefont{L.}~\bibnamefont{Glaser}},
  \bibinfo{author}{\bibfnamefont{P.~H.} \bibnamefont{Dederichs}},
  \bibinfo{author}{\bibfnamefont{M.}~\bibnamefont{Martins}},
  \bibinfo{author}{\bibfnamefont{S.}~\bibnamefont{Bl\"ugel}}, \bibnamefont{and}
  \bibinfo{author}{\bibfnamefont{W.}~\bibnamefont{Wurth}},
  \eprint{cond-mat/0608048}.

\bibitem[{\citenamefont{Bergman et~al.}(2006)\citenamefont{Bergman,
  Nordstr\"om, Klautau, Frota-Pess\^oa, and Eriksson}}]{Bergman06}
\bibinfo{author}{\bibfnamefont{A.}~\bibnamefont{Bergman}},
  \bibinfo{author}{\bibfnamefont{L.}~\bibnamefont{Nordstr\"om}},
  \bibinfo{author}{\bibfnamefont{A.~B.}~\bibnamefont{Klautau}},
  \bibinfo{author}{\bibfnamefont{S.}~\bibnamefont{Frota-Pess\^oa}},
  \bibnamefont{and} \bibinfo{author}{\bibfnamefont{O.}~\bibnamefont{Eriksson}},
  \bibinfo{journal}{Phys. Rev. B} \textbf{\bibinfo{volume}{73}},
  \bibinfo{pages}{174434} (\bibinfo{year}{2006}).

\bibitem[{\citenamefont{Andersen}(1975)}]{andersen75}
\bibinfo{author}{\bibfnamefont{O.~K.} \bibnamefont{Andersen}},
  \bibinfo{journal}{Phys. Rev. B} \textbf{\bibinfo{volume}{12}},
  \bibinfo{pages}{3060} (\bibinfo{year}{1975}).

\bibitem[{\citenamefont{Haydock}(1980)}]{haydock92}
\bibinfo{author}{\bibfnamefont{R.}~\bibnamefont{Haydock}},
  \emph{\bibinfo{title}{Solid State Physics}} (\bibinfo{publisher}{Academic},
  \bibinfo{address}{New York}, \bibinfo{year}{1980}),
  vol.~\bibinfo{volume}{35}, p. \bibinfo{pages}{216}.

\bibitem[{\citenamefont{Frota-Pess\^oa}(1992)}]{Frota92}
\bibinfo{author}{\bibfnamefont{S.}~\bibnamefont{Frota-Pess\^oa}},
  \bibinfo{journal}{Phys. Rev. B} \textbf{\bibinfo{volume}{46}},
  \bibinfo{pages}{14570} (\bibinfo{year}{1992}).

\bibitem[{\citenamefont{Klautau and Frota-Pess\^oa}(2005)}]{Klautau05}
\bibinfo{author}{\bibfnamefont{A.}~\bibnamefont{Klautau}} \bibnamefont{and}
  \bibinfo{author}{\bibfnamefont{S.}~\bibnamefont{Frota-Pess\^oa}},
  \bibinfo{journal}{Surf. Sci.} \textbf{\bibinfo{volume}{579}},
  \bibinfo{pages}{27} (\bibinfo{year}{2005}).

\bibitem[{\citenamefont{von Barth and Hedin}(1972)}]{barth72}
\bibinfo{author}{\bibfnamefont{V.}~\bibnamefont{von Barth}} \bibnamefont{and}
  \bibinfo{author}{\bibfnamefont{L.}~\bibnamefont{Hedin}}, \bibinfo{journal}{J.
  Phys. Chem.} \textbf{\bibinfo{volume}{5}}, \bibinfo{pages}{1629}
  (\bibinfo{year}{1972}).

\bibitem[{\citenamefont{K\"ubler et~al.}(1988)\citenamefont{K\"ubler, H\"ock,
  Sticht, and Williams}}]{Kubler}
\bibinfo{author}{\bibfnamefont{J.}~\bibnamefont{K\"ubler}},
  \bibinfo{author}{\bibfnamefont{K.-H.} \bibnamefont{H\"ock}},
  \bibinfo{author}{\bibfnamefont{J.}~\bibnamefont{Sticht}}, \bibnamefont{and}
  \bibinfo{author}{\bibfnamefont{A.~R.} \bibnamefont{Williams}},
  \bibinfo{journal}{J. Phys. F} \textbf{\bibinfo{volume}{18}},
  \bibinfo{pages}{469} (\bibinfo{year}{1988}).

\bibitem[{\citenamefont{Sandratskii and Guletskii}(1986)}]{Sandratskii}
\bibinfo{author}{\bibfnamefont{L.~M.} \bibnamefont{Sandratskii}}
  \bibnamefont{and} \bibinfo{author}{\bibfnamefont{P.~G.}
  \bibnamefont{Guletskii}}, \bibinfo{journal}{J. Phys. F}
  \textbf{\bibinfo{volume}{16}}, \bibinfo{pages}{L43} (\bibinfo{year}{1986}).

\bibitem[{\citenamefont{Nordstr\"om and Singh}(1996)}]{nord96}
\bibinfo{author}{\bibfnamefont{L.}~\bibnamefont{Nordstr\"om}} \bibnamefont{and}
  \bibinfo{author}{\bibfnamefont{D.~J.} \bibnamefont{Singh}},
  \bibinfo{journal}{Phys. Rev. Lett.} \textbf{\bibinfo{volume}{76}},
  \bibinfo{pages}{4420} (\bibinfo{year}{1996}).

\bibitem[{\citenamefont{Sandratskii}(1998)}]{Sandratskii2}
\bibinfo{author}{\bibfnamefont{L.}~\bibnamefont{Sandratskii}},
  \bibinfo{journal}{Adv. Phys.} \textbf{\bibinfo{volume}{91}},
  \bibinfo{pages}{47} (\bibinfo{year}{1998}).

\bibitem[{\citenamefont{Petrilli and Frota-Pess\^oa}(1990)}]{Petrilli90}
\bibinfo{author}{\bibfnamefont{H.~M.} \bibnamefont{Petrilli}} \bibnamefont{and}
  \bibinfo{author}{\bibfnamefont{S.}~\bibnamefont{Frota-Pess\^oa}},
  \bibinfo{journal}{J. Phys. Condens. Matter} \textbf{\bibinfo{volume}{2}},
  \bibinfo{pages}{135} (\bibinfo{year}{1990}).

\bibitem[{\citenamefont{Saha and Mookerjee}(2005)}]{Saha05}
\bibinfo{author}{\bibfnamefont{K.~K.} \bibnamefont{Saha}} \bibnamefont{and}
  \bibinfo{author}{\bibfnamefont{A.}~\bibnamefont{Mookerjee}},
  \bibinfo{journal}{J. Phys. Condens. Matter} \textbf{\bibinfo{volume}{17}},
  \bibinfo{pages}{287} (\bibinfo{year}{2005}).

\bibitem[{\citenamefont{Pick et~al.}(2003)\citenamefont{Pick, Stepanyuk,
  Baranov, Hergert, and Bruno}}]{Pick2003}
\bibinfo{author}{\bibfnamefont{S.}~\bibnamefont{Pick}},
  \bibinfo{author}{\bibfnamefont{V.~S.} \bibnamefont{Stepanyuk}},
  \bibinfo{author}{\bibfnamefont{A.~N.} \bibnamefont{Baranov}},
  \bibinfo{author}{\bibfnamefont{W.}~\bibnamefont{Hergert}}, \bibnamefont{and}
  \bibinfo{author}{\bibfnamefont{P.}~\bibnamefont{Bruno}},
  \bibinfo{journal}{Phys. Rev. B} \textbf{\bibinfo{volume}{68}},
  \bibinfo{eid}{104410} (\bibinfo{year}{2003}).

\bibitem[{\citenamefont{Beer and Pettifor}(1984)}]{Beer84}
\bibinfo{author}{\bibfnamefont{N.}~\bibnamefont{Beer}} \bibnamefont{and}
  \bibinfo{author}{\bibfnamefont{D.}~\bibnamefont{Pettifor}},
  \emph{\bibinfo{title}{The Electronic Structure of Complex Systems}}
  (\bibinfo{publisher}{Plenum Press}, \bibinfo{address}{New York},
  \bibinfo{year}{1984}).

\bibitem[{\citenamefont{Liechtenstein et~al.}(1987)\citenamefont{Liechtenstein,
  Katsnelson, Antropov, and Gubanov}}]{Liechtenstein87}
\bibinfo{author}{\bibfnamefont{A.~I.} \bibnamefont{Liechtenstein}},
  \bibinfo{author}{\bibfnamefont{M.~I.} \bibnamefont{Katsnelson}},
  \bibinfo{author}{\bibfnamefont{V.~P.} \bibnamefont{Antropov}},
  \bibnamefont{and} \bibinfo{author}{\bibfnamefont{V.~A.}
  \bibnamefont{Gubanov}}, \bibinfo{journal}{J. Magn. Magn. Matter}
  \textbf{\bibinfo{volume}{67}}, \bibinfo{pages}{65} (\bibinfo{year}{1987}).

\bibitem[{\citenamefont{Frota-Pess\^oa
  et~al.}(2000)\citenamefont{Frota-Pess\^oa, Muniz, and
  Kudrnovsky}}]{Frota-Pessoa00}
\bibinfo{author}{\bibfnamefont{S.}~\bibnamefont{Frota-Pess\^oa}},
  \bibinfo{author}{\bibfnamefont{R.~B.} \bibnamefont{Muniz}}, \bibnamefont{and}
  \bibinfo{author}{\bibfnamefont{J.}~\bibnamefont{Kudrnovsky}},
  \bibinfo{journal}{Phys. Rev. B} \textbf{\bibinfo{volume}{62}},
  \bibinfo{pages}{5293} (\bibinfo{year}{2000}).

\bibitem[{\citenamefont{Frota-Pess\^oa and Klautau}(2006)}]{soniaangela}
\bibinfo{author}{\bibfnamefont{S.}~\bibnamefont{Frota-Pess\^oa}}
  \bibnamefont{and} \bibinfo{author}{\bibfnamefont{A.}~\bibnamefont{Klautau}},
  \bibinfo{journal}{Int. J. Mod. Phys. B} \textbf{\bibinfo{volume}{20}},
  \bibinfo{pages}{5281} (\bibinfo{year}{2006}).

\bibitem[{\citenamefont{Tsoulos and Lagaris}(2006)}]{genprice}
\bibinfo{author}{\bibfnamefont{I.~G.} \bibnamefont{Tsoulos}} \bibnamefont{and}
  \bibinfo{author}{\bibfnamefont{I.~E.} \bibnamefont{Lagaris}},
  \bibinfo{journal}{Comp. Phys. Comm.} \textbf{\bibinfo{volume}{174}},
  \bibinfo{pages}{152} (\bibinfo{year}{2006}).

\bibitem[{\citenamefont{Ujfalussy et~al.}(1996)\citenamefont{Ujfalussy,
  Szunyogh, and Weinberger}}]{Ujfalussy96}
\bibinfo{author}{\bibfnamefont{B.}~\bibnamefont{Ujfalussy}},
  \bibinfo{author}{\bibfnamefont{L.}~\bibnamefont{Szunyogh}}, \bibnamefont{and}
  \bibinfo{author}{\bibfnamefont{P.}~\bibnamefont{Weinberger}},
  \bibinfo{journal}{Phys. Rev. B} \textbf{\bibinfo{volume}{54}},
  \bibinfo{pages}{9883} (\bibinfo{year}{1996}).

\bibitem[{\citenamefont{Lizarraga et~al.}(2004)\citenamefont{Lizarraga,
  Nordstr\"om, Bergqvist, Bergman, Sj\"ostedt, Mohn, and
  Eriksson}}]{Lizarraga04}
\bibinfo{author}{\bibfnamefont{R.}~\bibnamefont{Lizarraga}},
  \bibinfo{author}{\bibfnamefont{L.}~\bibnamefont{Nordstr\"om}},
  \bibinfo{author}{\bibfnamefont{L.}~\bibnamefont{Bergqvist}},
  \bibinfo{author}{\bibfnamefont{A.}~\bibnamefont{Bergman}},
  \bibinfo{author}{\bibfnamefont{E.}~\bibnamefont{Sj\"ostedt}},
  \bibinfo{author}{\bibfnamefont{P.}~\bibnamefont{Mohn}}, \bibnamefont{and}
  \bibinfo{author}{\bibfnamefont{O.}~\bibnamefont{Eriksson}},
  \bibinfo{journal}{Phys. Rev. Lett.} \textbf{\bibinfo{volume}{93}},
  \bibinfo{eid}{107205} (\bibinfo{year}{2004}).

\bibitem[{\citenamefont{P.~Mavropoulos and Bl\"ugel}(2006)}]{Mavropoulos06}
\bibinfo{author}{\bibfnamefont{R.~Z.} \bibnamefont{P.~Mavropoulos},
  \bibfnamefont{S.~Lounis}} \bibnamefont{and}
  \bibinfo{author}{\bibfnamefont{S.}~\bibnamefont{Bl\"ugel}},
  \bibinfo{journal}{Appl. Phys. A} \textbf{\bibinfo{volume}{82}},
  \bibinfo{pages}{103} (\bibinfo{year}{2006}).

\bibitem[{\citenamefont{Kr\"uger et~al.}(2000)\citenamefont{Kr\"uger, Taguchi,
  and Meza-Aguilar}}]{Kruger00}
\bibinfo{author}{\bibfnamefont{P.}~\bibnamefont{Kr\"uger}},
  \bibinfo{author}{\bibfnamefont{M.}~\bibnamefont{Taguchi}}, \bibnamefont{and}
  \bibinfo{author}{\bibfnamefont{S.}~\bibnamefont{Meza-Aguilar}},
  \bibinfo{journal}{Phys. Rev. B} \textbf{\bibinfo{volume}{61}},
  \bibinfo{pages}{15277} (\bibinfo{year}{2000}).

\bibitem[{\citenamefont{Spisak and Hafner}(2003)}]{Spisak03}
\bibinfo{author}{\bibfnamefont{D.}~\bibnamefont{Spisak}} \bibnamefont{and}
  \bibinfo{author}{\bibfnamefont{J.}~\bibnamefont{Hafner}},
  \bibinfo{journal}{Phys. Rev. B} \textbf{\bibinfo{volume}{67}},
  \bibinfo{eid}{134434} (\bibinfo{year}{2003}).

\bibitem[{\citenamefont{Hjortstam et~al.}(1996)\citenamefont{Hjortstam, Trygg,
  Wills, Johansson, and Eriksson}}]{Hjortstam96}
\bibinfo{author}{\bibfnamefont{O.}~\bibnamefont{Hjortstam}},
  \bibinfo{author}{\bibfnamefont{J.}~\bibnamefont{Trygg}},
  \bibinfo{author}{\bibfnamefont{J.~M.} \bibnamefont{Wills}},
  \bibinfo{author}{\bibfnamefont{B.}~\bibnamefont{Johansson}},
  \bibnamefont{and} \bibinfo{author}{\bibfnamefont{O.}~\bibnamefont{Eriksson}},
  \bibinfo{journal}{Phys. Rev. B} \textbf{\bibinfo{volume}{53}},
  \bibinfo{pages}{9204} (\bibinfo{year}{1996}).

\bibitem[{\citenamefont{Fernando and Cooper}(1988)}]{Fernando88}
\bibinfo{author}{\bibfnamefont{G.~W.} \bibnamefont{Fernando}} \bibnamefont{and}
  \bibinfo{author}{\bibfnamefont{B.~R.} \bibnamefont{Cooper}},
  \bibinfo{journal}{Phys. Rev. B}  \textbf{\bibinfo{volume}{38}},
  \bibinfo{pages}{3016} (\bibinfo{year}{1988}).

\bibitem[{\citenamefont{Ujfalussy et~al.}(2004)\citenamefont{Ujfalussy,
  Lazarovits, Szunyogh, Stocks, and Weinberger}}]{Ujfalussy04}
\bibinfo{author}{\bibfnamefont{B.}~\bibnamefont{Ujfalussy}},
  \bibinfo{author}{\bibfnamefont{B.}~\bibnamefont{Lazarovits}},
  \bibinfo{author}{\bibfnamefont{L.}~\bibnamefont{Szunyogh}},
  \bibinfo{author}{\bibfnamefont{G.~M.} \bibnamefont{Stocks}},
  \bibnamefont{and}
  \bibinfo{author}{\bibfnamefont{P.}~\bibnamefont{Weinberger}},
  \bibinfo{journal}{Phys. Rev. B} \textbf{\bibinfo{volume}{70}},
  \bibinfo{eid}{100404(R)} (\bibinfo{year}{2004}).

\bibitem[{\citenamefont{Lee et~al.}(2004)\citenamefont{Lee, Ho, and
  Persson}}]{Lee04}
\bibinfo{author}{\bibfnamefont{H.~J.} \bibnamefont{Lee}},
  \bibinfo{author}{\bibfnamefont{W.}~\bibnamefont{Ho}}, \bibnamefont{and}
  \bibinfo{author}{\bibfnamefont{M.}~\bibnamefont{Persson}},
  \bibinfo{journal}{Phys. Rev. Lett.} \textbf{\bibinfo{volume}{92}},
  \bibinfo{eid}{186802} (\bibinfo{year}{2004}).

\bibitem[{\citenamefont{Gotsis et~al.}(2006)\citenamefont{Gotsis, Kioussis, and
  Papaconstantopoulos}}]{gotsis2006}
\bibinfo{author}{\bibfnamefont{H.~J.} \bibnamefont{Gotsis}},
  \bibinfo{author}{\bibfnamefont{N.}~\bibnamefont{Kioussis}}, \bibnamefont{and}
  \bibinfo{author}{\bibfnamefont{D.~A.} \bibnamefont{Papaconstantopoulos}},
  \bibinfo{journal}{Phys. Rev. B} \textbf{\bibinfo{volume}{73}},
  \bibinfo{eid}{014436} (\bibinfo{year}{2006}).

\bibitem[{\citenamefont{Bergman et~al.}(2007)\citenamefont{Bergman,
  Nordstr\"om, Klautau, Frota-Pess\^oa, and Eriksson}}]{bergmanCr}
\bibinfo{author}{\bibfnamefont{A.}~\bibnamefont{Bergman}},
  \bibinfo{author}{\bibfnamefont{L.}~\bibnamefont{Nordstr\"om}},
  \bibinfo{author}{\bibfnamefont{A.}~\bibnamefont{Klautau}},
  \bibinfo{author}{\bibfnamefont{S.}~\bibnamefont{Frota-Pess\^oa}},
  \bibnamefont{and} \bibinfo{author}{\bibfnamefont{O.}~\bibnamefont{Eriksson}},
  \bibinfo{journal}{J. Phys. Condens. Matter} \textbf{\bibinfo{volume}{19}},
  \bibinfo{pages}{156226} (\bibinfo{year}{2007}).

\bibitem[{\citenamefont{A.~T.~Costa et~al.}(2005)\citenamefont{A.~T.~Costa,
  Muniz, and Mills}}]{Costa04}
\bibinfo{author}{\bibfnamefont{A.~T.}~\bibnamefont{Costa}},
  \bibinfo{author}{\bibfnamefont{R.~B.} \bibnamefont{Muniz}}, \bibnamefont{and}
  \bibinfo{author}{\bibfnamefont{D.~L.} \bibnamefont{Mills}},
  \bibinfo{journal}{Phys. Rev. Lett.} \textbf{\bibinfo{volume}{94}},
  \bibinfo{pages}{137203} (\bibinfo{year}{2005}).

\end{thebibliography}
\end{document}